\documentclass[preprint,12pt,authoryear]{elsarticle} 
\usepackage{lineno}
\usepackage{amsmath,amssymb,amsfonts,amsthm,latexsym,bm,hyphenat,tikz}
\usetikzlibrary{shapes,arrows}
\usetikzlibrary{decorations.pathreplacing,calligraphy}
\usepackage{tabularx,float,natbib,multirow,placeins,adjustbox}
\PassOptionsToPackage{hyphens}{url}
\usepackage{hyperref}
\DeclareMathOperator{\E}{\mathbb{E}}
\DeclareMathOperator*{\argmax}{argmax}
\newcounter{para}

\usepackage{setspace}
\usepackage[margin=2.5cm]{geometry}

\makeatletter
\def\ps@pprintTitle{%
  \let\@oddhead\@empty
  \let\@evenhead\@empty
  \let\@oddfoot\@empty
  \let\@evenfoot\@oddfoot
}
\makeatother

\begin{document}
\hfuzz=50pt
\hbadness=20000
\sloppy

\begin{frontmatter}
\title{Beyond expected values: Making environmental decisions using value of information analysis when measurement outcome matters}
\author[QUT,CDS,SAEF]{Morenikeji D. Akinlotan} \ead{morenikeji.akinlotan@qut.edu.au}
\author[QUT,CDS]{David J. Warne} \ead{david.warne@qut.edu.au}
\author[QUT,CDS,SAEF]{Kate J. Helmstedt} \ead{kate.helmstedt@qut.edu.au}
\author[QUT,CDS]{Sarah A. Vollert} \ead{sarah.vollert@connect.qut.edu.au}
\author[CSIRO_Iadine]{Iadine Chad\`{e}s} \ead{iadine.chades@csiro.au}
\author[QUT,CDS,USC]{Ryan F. Heneghan} \ead{rheneghan@usc.edu.au}
\author[Deakin,CSIRO_Hui]{Hui Xiao} \ead{hui.xiao@csiro.au}
\author[QUT,CDS,SAEF,SCC]{Matthew P. Adams\corref{cor1}} \ead{mp.adams@qut.edu.au}
\cortext[cor1]{Corresponding Author}
\address[QUT]{School of Mathematical Sciences, Queensland University of Technology, Brisbane QLD 4001, Australia}
\address[CDS]{Centre for Data Science, Queensland University of Technology, Brisbane QLD 4001, Australia}
\address[SAEF]{Securing Antarctica's Environmental Future, Queensland University of Technology, Brisbane QLD 4001, Australia}
\address[CSIRO_Iadine]{Land and Water Flagship, Commonwealth Scientific and Industrial Research Organisation, Dutton Park QLD 4102, Australia}
\address[USC]{School of Science, Technology and Engineering, University of the Sunshine Coast, Petrie QLD 4502, Australia}
\address[Deakin]{School of Life and Environmental Sciences, Deakin University, Burwood VIC 3125, Australia}
\address[CSIRO_Hui]{Commonwealth Scientific and Industrial Research Organisation, Queensland Biosciences Precinct, St Lucia QLD 4067, Australia}
\address[SCC]{School of Chemical Engineering, The University of Queensland, St Lucia QLD 4072, Australia}

\begin{abstract}
In ecological and environmental contexts, management actions must sometimes be chosen urgently. Value of information (VoI) analysis provides a quantitative toolkit for projecting the improved management outcomes expected after making additional measurements. However, traditional VoI analysis reports metrics as expected values (i.e.\ risk-neutral). This can be problematic because expected values hide uncertainties in projections. The true value of a measurement will only be known after the measurement's outcome is known, leaving large uncertainty in the measurement's value before it is performed. As a result, the expected value metrics produced in traditional VoI analysis may not align with the priorities of a risk-averse decision-maker who wants to avoid low-value measurement outcomes. In the present work, we introduce four new VoI metrics that can address a decision-maker's risk-aversion to different measurement outcomes. We demonstrate the benefits of the new metrics with two ecological case studies for which traditional VoI analysis has been previously applied. In the first case study concerning a test for disease presence at a potential frog translocation site, traditional VoI analysis predicts the test yields an additional expected gain of approximately 10 frogs. However, our new VoI metrics also highlight a 40\% risk that the test is valueless; this knowledge may deter a risk-averse decision-maker from doing the test. In the second case study concerning the design of a trial release prior to a large-scale turtle reintroduction, traditional and new VoI metrics have consistent predictions of which design to choose. However, whilst the best trial release design will increase expected turtle survival in the wild by only 3\%, the new VoI metrics find that this trial design has a 94\% probability of improving the design of the large-scale turtle reintroduction. Using the new metrics, we also demonstrate a clear mathematical link between the often-separated environmental decision-making disciplines of VoI and optimal design of experiments. This mathematical link has the potential to catalyse future collaborations between ecologists and statisticians to work together to quantitatively address environmental decision-making questions of fundamental importance. Overall, the introduced VoI metrics complement existing metrics to provide decision-makers with a comprehensive view of the value of, and risks associated with, a proposed monitoring or measurement activity. This is critical for improved environmental outcomes when decisions must be urgently made.\\

\end{abstract}
\end{frontmatter}

\vspace{0.2cm}
\section{Introduction} \label{sec:Intro}
\noindent In many fields, including ecology, environmental monitoring, economics and health, decision-makers must identify whether there is sufficient information to confidently choose management actions, or if further information should be acquired before an action is chosen \citep{Bennett2018,Yokota2004}. Ideally, the expected benefit of additional measurements outweighs the cost \citep{Koerkamp2006}, but in ecological and environmental contexts, management actions must sometimes be chosen urgently \citep{McDonaldMadden2010}. Management actions always carry a risk of unanticipated and detrimental outcomes \citep{Bergstrom2009}, but delays in management actions can be extraordinarily costly, resulting in ecosystem damage or potential species extinctions \citep{Martin2012}. Thus, decision-makers may need to quickly decide if more information should be obtained about the system of interest before choosing an action, or if decisions should be promptly made based on current knowledge \citep{Bennett2018}. If a timed sequence of decisions about management actions is to account for the system's change following each action, this sequential decision-making process can be informed using mathematical approaches that explicitly describe the iterative interaction between management actions and system changes \citep{Chades2008,Chades2021}. However, even if an immediate decision about which \textit{one} management action to choose at \textit{one} specific time would benefit from additional information about the system, conservation monitoring is very costly and time consuming. Thus, it is crucial that such monitoring is justified by its usefulness for improving management decisions \citep{McDonaldMadden2010}. 

Value of information (VoI) analysis provides a quantitative toolkit for assessing the expected improvement that additional information can make to management decisions (\citealt{Canessa2015}; \citealt{Jackson2022}). VoI represents the difference between expected outcomes when a management decision is made in the presence of new information versus only current information \citep{Yokota2004}. VoI analysis therefore addresses whether or not it is beneficial to collect additional information when improved decision-making is the primary objective. When the management action to be chosen must be performed at one specific time, then VoI analysis precedes both the (proposed) information collection activity and the choice of management action (Figure \ref{fig:timeline}).

\tikzstyle{line} = [draw, very thick, -latex']
\tikzstyle{flow} = [line, line width=0.4mm]
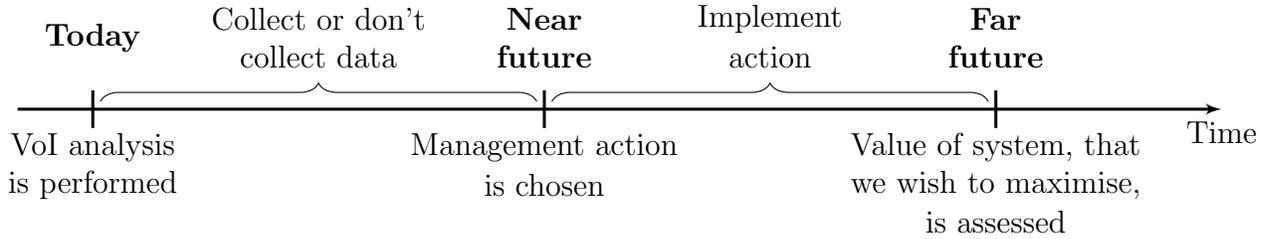
\begin{figure}[ht!]\centering
\begin{tikzpicture}
\node (tstart) at (0,0) {};
\node (tend) at (16,0) {};
\path [flow] (tstart.center) -- node [at end,below] {Time} (tend.center);
\node (todayup) at (1,0.25) {};
\node (todaydown) at (1,-0.25) {};
\draw [line width=0.4mm] (todayup.center) -- (todaydown.center);
\node (todaylabel) at (1,0.95) {\textbf{Today}};
\node (todaylabel2) at (1,-0.5) {VoI analysis};
\node (todaylabel3) at (1,-1) {is performed};
\draw [decorate,decoration = {brace,amplitude=7pt}] (1.1,0.1) --  (6.9,0.1);
\node (collectdata) at (4,1.2) {Collect or don't};
\node (collectdata2) at (4,0.7) {collect data};
\node (decisionup) at (7,0.25) {};
\node (decisiondown) at (7,-0.25) {};
\draw [line width=0.4mm] (decisionup.center) -- (decisiondown.center);
\node (decisionlabel) at (7,1.2) {\textbf{Near}};
\node (decisionlabel) at (7,0.7) {\textbf{future}};
\node (decisionlabel2) at (7,-0.5) {Management action};
\node (decisionlabel3) at (7,-1) {is chosen};
\draw [decorate,decoration = {brace,amplitude=7pt}] (7.1,0.1) --  (12.9,0.1);
\node (implementaction) at (10,1.2) {Implement};
\node (implementaction2) at (10,0.7) {action};
\node (resultup) at (13,0.25) {};
\node (resultdown) at (13,-0.25) {};
\draw [line width=0.4mm] (resultup.center) -- (resultdown.center);
\node (resultlabel) at (13,1.2) {\textbf{Far}};
\node (resultlabel) at (13,0.7) {\textbf{future}};
\node (resultlabel2) at (13,-0.5) {Value of system, that};
\node (resultlabel3) at (13,-1) {we wish to maximise,};
\node (resultlabel4) at (13,-1.5) {is assessed};
\end{tikzpicture}
\caption{Overview of the timing of value of information (VoI) analysis, for a decision problem where a management action needs to be chosen at some `near future' time, for a system whose value we want to maximise in the `far future'. In this system, there is a measurement (`collect data') that could be performed in the meantime to improve the action that is chosen. VoI analysis informs whether this measurement should be performed or not. In this paper we focus on decision problems which adhere to this timeline.}
\label{fig:timeline}
\end{figure}

Traditional calculations in VoI analysis assume that the decision-maker is risk-neutral. This is not realistic in many management contexts, especially in conservation where it is expected that many decision-makers responsible for environmental management are averse to risk \citep{Tulloch2015}. A decision-maker's aversion to risk can be incorporated into VoI analysis calculations \citep{Xiao2019}, and previous work has focused on reducing the risk of undesirable outcomes for the system as a whole \citep{Nadiminti1996,Borgonovo2017}. For example, aversion to the risk of poor outcomes can be incorporated in VoI analysis with a strategically-chosen nonlinear value function penalising low values of the system's properties that the decision-maker aims to maximise (e.g.\ \citealt{Hilton1981,Willinger1989,Bleichrodt2020}).

A similarly important concern is that the information obtained from a measurement does not substantially improve the decision-maker's choice of management action. For example, if new information does not change the action the decision-maker was otherwise going to select, the data collection activity may be deemed retrospectively as ``worthless''. Whilst this potential outcome is not directly detrimental to the system being managed, the possibility of this outcome may be perceived by the decision-maker as a ``risk'' of wasted investment of time and/or money in monitoring.

More broadly, there may be a range of potential measurement outcomes which are viewed as ``low-value measurement outcomes'' -- they may change the decision-maker's chosen management action, but due to their expected low improvement values they are deemed by the decision-maker to be not worth the cost of measurement or delay in action \citep{Grantham2009}. Even if traditional VoI analysis suggests that a new measurement has a substantial \textit{expected} value, an expected value says nothing about the \textit{uncertainty} in this value, so there may still be an unacceptably high probability that the measurement yields low value to the decision-maker. Whilst it may be impossible to know if a low-value measurement outcome will definitely occur prior to the data being collected, the \textit{risk} of low-value measurement outcomes, on the other hand, can be predicted, and to our best knowledge the calculation of this risk has not yet been explored. Such risks are particularly critical to quantify in environmental decision-making when the consequences of a delayed management action can be catastrophic \citep{Martin2012}.

Due to its focus on informing data collection activities, VoI is also related to optimal design \citep{Williams2020}, a class of experimental design methods that aims to recommend monitoring activities based on some statistical optimality criteria \citep{Holling2017}. VoI analysis has evolved primarily from the economics literature \citep{Raiffa1968}, with subsequent uptake of this analysis in medical decision-making \citep{Felli1998} and increasingly recent uptake by the ecological and environmental science communities (\citealt{Runge2011}; \citealt{Nicol2019}; \citealt{Lawson2022}; \citealt{Stantial2023}). The research area of optimal design has been expanding rapidly in parallel in the statistics literature (\citealt{Ryan2016}; \citealt{Blau2022}; \citealt{Warne2022}). However, the topics of VoI and optimal design have not substantially overlapped thus far in the literature due to these topics' different origins in economics and statistics respectively. The \textit{conceptual} link between VoI analysis and optimal design is already clear, as both are decision-theoretic approaches that estimate the expected benefits of collecting further information \citep{Jackson2022}. Making the \textit{mathematical} link between VoI analysis and optimal design clear has the potential to unlock the rapid technical advances made by statisticians in the field of optimal design to help address VoI questions in ecology and environmental science.

Here, we extend VoI analysis by introducing four new metrics that explicitly address risk tolerance to different measurement outcomes. In doing so we also show how the new metrics can be used to clearly establish the mathematical link between VoI and optimal design. Our new metrics quantify (Table \ref{table:new_metrics}):
\begin{enumerate}
\item The shift $\Delta$EV$_x$ in the expected value of the system, dependent on measurement outcome $x$;
\item The value VSI$_x$ of sample information, dependent on measurement outcome (this value of the \textit{information} can be different to the \textit{shift in expected value of the system} resulting from collecting this information);
\item The uncertainty (specifically, standard deviation $\sigma$VSI) in the value of information provided by the measurement; and
\item The risk rVSI$_\delta$ of the measurement providing low value (specifically, probability of value $\leq \delta$) for decision-making.
\end{enumerate}
When reported alongside traditional VoI metrics, these new metrics provide the decision-maker with a broader perspective on the potential value of proposed measurements before choosing whether (and what) information should be collected. For example, the decision-maker would be able to choose monitoring activities that either maximise the expected value of the monitoring or minimise the risk of low-value monitoring outcomes. Finally, we demonstrate the usefulness of our new metrics via two ecological case studies that traditional VoI metrics have been previously applied to \citep{Canessa2015}. Using these case studies, we show how our new metrics help decision-makers to form a comprehensive view of the value of, and risks associated with, a proposed monitoring or measurement activity; this is critical for improved environmental outcomes when decisions must be urgently made.

\nohyphens{
\renewcommand{\arraystretch}{1.3}
\begin{table}[!ht]
\centering \scriptsize
\caption{Summary of the four new metrics proposed in this work. The `example' column is drawn from Case Study 1 (Section \ref{sec:Case1}).}
\label{table:new_metrics}
\begin{tabular}{c p{0.15\linewidth} p{0.35\linewidth} p{0.35\linewidth}}
\\
\hline
\textbf{Metric} & \textbf{Definition} & \textbf{Interpretation} & \textbf{Example} \\ \hline
$\Delta \mathrm{EV}_x$ & Shift in expected value of the system (Section \ref{sec:Delta_EV_x}) & Given an observed new data point $x$, how much has this changed our belief of the system's expected value? & If a test for disease in frogs is negative, the expected frog population increases by $\approx$17 frogs. \\
$\mathrm{VSI}_x$ & Value of sample information (Section \ref{sec:VSIx}) & Given an observed new data point $x$, how much is the system's updated expected value improved by choosing a better action (if one exists) compared to the system's updated expected value if we still chose to carry out our original action? & If a test for disease in frogs is negative, this changes our management action, which results in an increase in the expected frog population by $\approx$17 frogs. \\
$\sigma \mathrm{VSI}$ & Standard deviation of the value of sample information (Section \ref{sec:sigma_VSI}) & What is the uncertainty in the value of the measurement, before knowing what new data point $x$ is obtained? & By obtaining new information, the management action may or may not change, depending on the content of the information. However, overall, the frog population is expected to increase by 10.4$\pm$8.4 frogs. (Here, $\sigma$VSI$=$8.4, and the traditional metric EVSI$=$10.4.) \\
$\mathrm{rVSI}_\delta$ & Risk of low value of sample information (Section \ref{sec:rVSI_delta}) & What is the probability that the measurement will have low value for decision-making, before knowing what new data point $x$ is obtained? (This metric may be of particular relevance to risk-averse decision-makers.) & There is a 40\% risk that the measurement's impact on decision-making is no change in the expected frog population.\\
\hline
\end{tabular}
\end{table}
\renewcommand{\arraystretch}{1}}

\section{Materials and methods} \label{sec:method}

\noindent To provide context for our new VoI analysis metrics (Table \ref{table:new_metrics}), we first introduce the traditional metrics used in VoI analysis (Section \ref{sec:def}) and then explain each of the four new metrics in turn in Sections \ref{sec:Delta_EV_x}-\ref{sec:rVSI_delta}. We then mathematically describe the link between VoI and optimal design (Section \ref{sec:optdesign}) as this link requires knowledge of our second new metric (VSI$_x$, introduced in Section \ref{sec:VSIx}). Finally, we apply our new metrics to two case studies (Sections \ref{sec:Case1} and \ref{sec:Case2}), to compare and contrast the insights obtained from the traditional and new VoI metrics for environmental decision-making.

\subsection{Common calculations and terminology in VoI analysis}  \label{sec:def}
\noindent To perform VoI analysis, one must first describe how the \textit{value} $V(a,s)$ of the system being managed depends on the system's \textit{state} $s$ and the management \textit{action} $a$ being undertaken \citep{Canessa2015}. Values \(V(a,s)\) are quantified in units of the management objective, for example, population size or the probability of species survival. It is assumed that the decision-maker wishes to maximise the system's value. The current uncertainty about the system can be represented as alternative potential states \(s\) of the system, with each state assigned a prior probability \(p(s)\) of being true. If the management action \(a\) is performed on the system without collecting any new data prior to this action, the action is said to be taken ``under uncertainty''. An example of such uncertainty in a system is the presence or absence of a species when no new information is obtained to confirm species presence or absence \citep{Bennett2018}.  

The expected value \(\text{EV}(a)\) of the system if an action \(a\) is taken under uncertainty is calculated from the expectation of the system's value over all possible states $s$, given that the action $a$ is definitely chosen:
	\begin{equation}
		\text{EV}(a) = \E_{s} [V(a,s)]. \label{eq:EV}
	\end{equation}
A risk-neutral decision-maker will choose the action that yields the largest expected value of the system. If this action is selected, the expected value of the system, denoted as EV$_{\text{uncertainty}}$, is
	\begin{equation} 
		\text{EV}_\text{uncertainty} = \max_a \{\text{EV}(a)\}.    \label{eq:EVuncert}
	\end{equation}
Conversely, if perfect information is obtained for the system state, a risk-neutral decision-maker can choose the action that yields the maximum value possible for that system state. Such a decision is said to be made ``under certainty''. In this case, the expected value of the system, denoted as EV$_{\text{certainty}}$, is
	\begin{equation} 
		\text{EV}_{\text{certainty}} = \E_{s} [\max_a \{ V(a,s) \} ].  \label{eq:EVcert}  
	\end{equation}
Equation \eqref{eq:EVcert} indicates that under each fully-known system state \(s\), the action \(a\) that yields the maximum value of \(V(a,s)\) is chosen by the decision-maker. However, the use of the expectation over all system states \(s\) in Equation~\eqref{eq:EVcert} acknowledges that this calculation is carried out before the perfect information is obtained (see Figure \ref{fig:timeline}). 

When performing VoI analysis, it is assumed that every decision-maker seeks to choose the action that leads to the greatest net benefit \citep{Yokota2004}. Thus, a first commonly-calculated VoI metric -- the expected value of perfect information (EVPI) -- is the expected benefit of resolving or eliminating uncertainty entirely. Mathematically, EVPI is the difference between the expected values of the system when the best management actions are taken under certainty versus under uncertainty: 
	\begin{align}
   		\text{EVPI} & = \text{EV}_{\text{certainty}} - \text{EV}_{\text{uncertainty}} \nonumber  \\ 
   		& = \E_{s} [\max_a \{ V(a,s) \} ] - \max_a \{ \E_{s} [V(a,s)] \}. \label{eq:EVPI}
	\end{align}
Since access to perfect information is usually not possible in practice, Equation~\eqref{eq:EVPI} represents a best-case scenario of information gain from measurements prior to action. The EVPI therefore puts an upper bound on the expected value of imperfect information, and as such is a useful quantity to estimate even when the proposed measurements are expected to yield information that is nowhere near perfect. For example, if a risk-neutral decision-maker deems the EVPI to be low enough, then no measurement should be performed before the management action is chosen. This outcome is likely to be common \citep{McDonaldMadden2010}, and can be advantageous because actions are not delayed by potentially costly measurements.

Measurements that do not yield perfect information can, however, still empower the decision-maker to choose better management actions \citep{Yokota2004}. Thus, a second commonly calculated VoI metric -- the expected value of sample information (EVSI) -- is the increase in the system's expected value when the decision-maker obtains a sample of imperfect measurements, observations, or information \citep{Canessa2015}. Calculating EVSI is slightly more complicated than EVPI; it requires an expression for the expected value of additional imperfect information, but this imperfect information, denoted as $x$, is not guaranteed to completely eliminate uncertainty regarding which state $s$ the system is in. For a risk-neutral decision-maker, the expected value of the system after obtaining imperfect information, denoted as \(\text{EV}_{\text{less uncertainty}}\), is
	\begin{equation}
		\text{EV}_{\text{less uncertainty}} = \E_{x} [ \max_a \{ \E_{s|x} [ V(a,s) ] \} ],  \label{eq:EVlessuncert}  
	\end{equation}
where \(x\) is the specific outcome obtained by the proposed measurement of imperfect information. For example, a monitoring activity to determine the number of koalas in a forest area may locate 10 koalas ($x=10$) when there are in fact 14 koalas present in the area ($s=14$). However, we do not know prior to monitoring how many koalas will actually be found ($x =\,\,?$), so calculating $\text{EV}_{\text{less uncertainty}}$ requires an expectation over all possible measurement outcomes $x$.

The EVSI is thereafter calculated as follows:
\begin{align}
\text{EVSI} & = \text{EV}_{\text{less uncertainty}} - \text{EV}_{\text{uncertainty}} \nonumber \\  
    & = \E_{x} [ \max_a \{ \E_{s|x} [V(a,s)] \} ] - \max_a \{\E_{s} [V(a,s)]\}. \label{eq:EVSI} 
\end{align}
The EVSI can be used by the decision-maker to decide whether or not to perform a measurement of imperfect information prior to choosing the management action: if the EVSI is judged to be sufficiently large, this provides greater impetus to do the measurement first.

However, EVSI is only appropriate if the decision-maker is indifferent to the potential for zero, very small, or very large improvements in the expected system state that may result from different  measurement outcomes. An \textit{expected} value of information, calculated before the measurement is performed (Figure \ref{fig:timeline}), does not provide any indication of the \textit{uncertainty} in the actual value of information that will be obtained from the measurement. To inform a decision-maker that wishes to avoid low-value measurement outcomes, new metrics are needed.

\subsection{New metrics to quantify risk and uncertainty in VoI analysis} \label{sec:VOIanalysis}
\noindent We propose four new metrics that describe the risk and uncertainty associated with different potential measurement outcomes. If the analysis of a decision problem explicitly accounts for the decision-maker's risk tolerance (e.g.\ risk-neutral or risk-averse), it is hoped that the outcomes of the analysis will be of greater practicality to the decision-maker \citep{Markowitz1959}. With this in mind, we assume that a risk-neutral decision-maker will aim for the highest expected outcome, whilst a risk-averse decision-maker will aim to minimise the risks or regrets associated with an undesirable outcome  \citep{Tulloch2015,Xiao2019}. This ``risk'' is equally applicable to a decision-maker's tolerance to low-value system outcomes \citep{Nadiminti1996,Borgonovo2017} and low-value measurement outcomes (addressed in the present work). 

The four new metrics are summarised in Table \ref{table:new_metrics}, and mathematically detailed in Sections \ref{sec:Delta_EV_x}-\ref{sec:rVSI_delta}. All four metrics are designed for the realistic situation in which the proposed measurement yields imperfect information, and thus the new metrics complement EVSI. However, although not discussed further here, the new metrics can also be applied to situations in which the proposed measurement yields perfect information (to complement EVPI), as this latter case is a mathematical simplification of imperfect information in which every measurement outcome $x$ reveals the true state $s$ of the system. Thus, any formulae introduced in this paper for the case of imperfect information can also be used for the case of perfect information, by mathematically substituting the system state $s$ into all instances of the measurement outcome $x$ (which subsequently eliminates $x$ from these formulae).

\subsubsection{Metric 1: Shift in expected value of the system, $\Delta \mathrm{EV}_x$}
\label{sec:Delta_EV_x}
\noindent The first new metric we introduce is the shift in expected system value due to the measurement, and is denoted by $\Delta \text{EV}_x$. This metric focuses on the value of the \textit{system} rather than the value of the \textit{information} associated with the measurement, although it has a clear and strong link to the traditional VoI metric of EVSI that makes it a logical new metric to discuss first. \(\Delta \text{EV}_x\) also aids with understanding the three other, more decision-focused, metrics we introduce in this work. 

The metric $\Delta \text{EV}_x$ is defined for a particular measurement outcome $x$, so every potential measurement outcome $x$ yields a different value of $\Delta \text{EV}_x$. This metric is obtained by removing the expectation over measurement outcomes \(x\) from the formula for EVSI (Equation~\eqref{eq:EVSI}) so that $\text{EVSI} = \E_x [ \Delta \text{EV}_x ]$. Since the second term in Equation~\eqref{eq:EVSI} has no dependence on measurement outcome $x$, any expectation over $x$ introduced to or removed from this term has no effect (see Supplementary Material Section S1.1 for further mathematical details). Hence, we obtain:
\begin{align}
  \Delta \text{EV}_x &= \max_a \{ \E_{s|x} [V(a,s)] \}  - \max_a \{ \E_{s} [V(a,s)] \} \nonumber \\
   &= \max_a \{ \E_{s|x} [V(a,s)] \}  - \text{EV}_{\text{uncertainty}}.  \label{eq:deltaEV}   
\end{align}
$\Delta \text{EV}_x$ differs from EVSI, which specifies the expected value of the \textit{measurement information} and is calculated \textit{without knowing} what the measurement outcome is. Instead, $\Delta \text{EV}_x$ specifies the change in the expected value of the \textit{system} if \textit{we know} what the measurement outcome is ($x$). Hence, \(\Delta \text{EV}_x\) estimates the shift in expected system value for a particular measurement outcome $x$, but not necessarily the shift in decision. Note that new knowledge gained from a measurement can lead to a re-evaluation of the system state to possess either a higher or lower expected value. So, unlike the strictly non-negative VoI metrics EVSI and EVPI, the metric \(\Delta \text{EV}_x\) can be positive, zero or negative. The system-focused metric \(\Delta \text{EV}_x\) is also a useful complement to the more decision-focused metric \( \text{VSI}_x \), which we introduce next.

\subsubsection{Metric 2: Value of sample information, $\mathrm{VSI}_x$}
\label{sec:VSIx}

\noindent The second new metric is the value of sample information (VSI). VSI is the value that imperfect information has for decision-making, specified \textit{after} the information is collected. Similarly to $\Delta \text{EV}_x$, the VSI is defined for a particular measurement outcome $x$, so is denoted mathematically as VSI$_x$. However, unlike $\Delta \text{EV}_x$, VSI$_x$ is a measure of information value rather than system value. Specifically, VSI$_x$ is the expected value of the \textit{measurement information} if \textit{we know} what the measurement outcome is ($x$).

To explain how VSI$_x$ is obtained, we first consider the action that yields the maximum expected value of the system if no measurement is performed. If we denote this optimal action under uncertainty as $a^*$, from Equations~\eqref{eq:EV} and \eqref{eq:EVuncert} it is clear that
\begin{equation}
    a^* =  \underset{a}{\argmax} \{ \E_{s} [V(a,s)] \}.  \label{eq:astar}
\end{equation}
Action \(a^*\) would be chosen by the decision-maker if no measurement is made.

Now, suppose that the measurement has taken place and a particular measurement outcome \(x\) is observed. We calculate VSI$_x$ associated with this measurement outcome \(x\) as the difference between the expected value of the system if the best action is taken after knowing the measurement outcome is \(x\) and the expected value of the system if the best action is taken without the knowledge that the measurement outcome will indeed be \(x\): 
\begin{equation}
\text{VSI}_{x} =  \max_a \{ \E_{s|x} [V(a,s)] \}  - \E_{s|x} [V(a^*,s)].  \label{eq:VSIx}
\end{equation}
Note that the dependence on $x$ of the expectation in the second term of Equation \eqref{eq:VSIx} can be interpreted in the following way: the VSI is calculated as if the measurement has already taken place and the decision-maker is reflecting on the expected value gained from the measurement in hindsight.

From the definition of VSI$_x$ provided in Equation~\eqref{eq:VSIx}, we also note that the expectation of \(\text{VSI}_{x}\) taken over all possible measurement outcomes \(x\) is equal to the EVSI defined in Equation~\eqref{eq:EVSI} (see Section S1.2 for further mathematical details). Hence, the expectations over $x$ of the two new metrics $\Delta \text{EV}_x$ and VSI$_x$ both yield EVSI, even though $\Delta \text{EV}_x$ and VSI$_x$ have different mathematical definitions. The metric $\Delta \text{EV}_x$ represents the shift in expected value of the system after obtaining the measurement outcome $x$, so $\Delta \text{EV}_x$ can be positive, zero or negative. On the other hand, VSI$_x$ represents the value for decision-making gained from the measurement outcome $x$, so VSI$_x$ is, like EVSI, strictly non-negative. We will see later in the case studies (Section \ref{sec:CaseStudies}) that sometimes $\Delta \text{EV}_x = \text{VSI}_x$, but this is not true in general.

While the two new metrics $\Delta \text{EV}_x$ and VSI$_x$ reveal far more information than the traditional EVSI metric alone, these two new metrics can also be unwieldy to report if there are many (or infinite) possible measurement outcomes $x$ that can be obtained. The next two metrics we introduce are not dependent on measurement outcome, so advantageously yield only one number each, although calculating both of these metrics \textit{requires} that VSI$_x$ is calculated first.

\subsubsection{Metric 3: Standard deviation of the value of sample information, $\sigma \mathrm{VSI}$}
\label{sec:sigma_VSI}
\noindent The traditional metric of EVSI is the \textit{expected} value of sample information, and thus provides no measure of the \textit{uncertainty} in the value of information that might actually be obtained. Hence, EVSI can be viewed as an ``average'' VoI, but it would be useful to be able to report VoI as ``average $\pm$ uncertainty''. Variance (the square of standard deviation) has long been suggested as a measure of risk when deciding which management actions to undertake \citep{Markowitz1991}, although standard deviation is easier to interpret because it is in the same units as the quantity being estimated. Hence, the third metric we introduce is the standard deviation in the value of sample information, denoted as \(\sigma \text{VSI}\): 
\begin{equation}
    \sigma \text{VSI} =  \sigma_x [ \text{VSI}_x ] = \sqrt{\text{Var}_x [\text{VSI}_x]}, \label{eq:SDVSI}
\end{equation}
where \(\text{Var}_x[\text{VSI}_x]\) is the variance in the distribution of \(\text{VSI}_x\) across all different measurement outcomes \(x\). If \(\sigma \text{VSI}\) is reported alongside EVSI, then these two metrics together characterise the mean and standard deviation of VoI across all potential outcomes $x$ of a proposed imperfect measurement, i.e.\ EVSI $\pm$ $\sigma$VSI.

\subsubsection{Metric 4: Risk of low value of sample information, $\mathrm{rVSI}_\delta$}
\label{sec:rVSI_delta}
\noindent The fourth and final metric that we introduce quantifies the risk that the measurement will yield low value for decision-making. For this metric, a threshold value of VSI is chosen (ideally by the decision-maker) at or below which the value of the measurement is considered ``too low''. This threshold value of VSI is denoted as $\delta$ and reflects the decision-maker's risk aversion to low-value measurement outcomes. The risk of low value of sample information, denoted as \(\text{rVSI}_\delta\), is  defined as the probability across all measurement outcomes $x$ that VSI$_x \leq \delta$:
\begin{equation}
   \text{rVSI}_\delta = p(\text{VSI}_x \leq \delta). \label{eq:riskVSI}
\end{equation}
Here, \(\delta\) can be interpreted as a measure of the decision-maker's level of risk tolerance, because it defines the minimum value of the measurement above which they are willing to accept.  For example, \(\text{rVSI}_{10} = 0.15\) indicates that the decision-maker desires that the measurement's expected value should exceed ten, and the risk that this desire is not achieved is 15\%. To obtain rVSI$_{10}$, we would first identify all measurement outcomes $x$ that yield VSI$_x \leq 10$, and calculate the probabilities $p(x)$ of each of these outcomes. Adding together these probabilities $p(x)$ would yield rVSI$_{10}$. If the decision-maker considers this risk to be acceptable, they may wish to proceed with the proposed measurement prior to choosing a management action. 
 
The threshold value $\delta$ must be non-negative. If it is set to zero, then rVSI$_0$ is the probability (i.e.\ risk) that the measurement has no value because its outcome did not change the management action chosen by the decision-maker. If an obtained measurement outcome suggests that (1) the originally proposed management action is still the best and (2) this best action is expected to have an even greater positive effect on the system than previously thought, the value of the measurement for decision-making is still zero because it did not change the decision-maker's action. 

\subsection{The mathematical link between VoI and optimal design} \label{sec:optdesign}
\noindent Our new metrics allow us to establish the mathematical connection between the research fields of optimal design and VoI theory. It is our hope that establishing this theoretical link will encourage advances made in either of these two fields to advance the other field as well. Briefly, optimal design is a class of experimental design based on optimising certain statistical criteria in relation to a prescribed task such as parameter estimation \citep{Holling2017}. Experimental design provides the theoretical framework for allocating resources when the design of an experiment or gathering more information about an uncertain system state would enable the decision-maker to better understand an underlying process \citep{Raiffa1968,Ryan2016}.

Optimal design minimises the experimental resources required to make precise inferences on the relevant features of an experimental design \citep{Ryan2016,Holling2017}. Thus, optimal design has broad application in many fields, including the medical, social, natural, and educational sciences, and in business, finance and engineering \citep{Drovandi2013,Gill2022}. With the advent of increasing computational power, optimal design utilises Bayesian approaches to incorporate prior information and/or the relevant system uncertainties within a \textit{utility function} that describes the aim of the experiment \citep{Ryan2016}. 

In statistical experimental design, there may be some uncertainty about the parameters controlling the system behaviour. Accounting for this lack of knowledge (of the model parameters or the model itself) can help to design  efficient experiments. In  Bayesian optimal designs, current knowledge (prior to any proposed experiments) of system parameter estimates can be represented by the prior distribution $p(\theta)$ \citep{Drovandi2013}. This prior distribution can incorporate information or data from previous experimental studies, expert-elicited data, and/or subjective belief of the experimenters \citep{Ryan2016,Drovandi2013}. Thus, Bayesian optimal designs are designs that account for pre-existing knowledge of uncertainty in parameter estimates. Following the decision-theoretical approach presented by \cite{Lindley1972}, it involves the use of a design criterion or utility function $u(d,x,\theta)$, where $x$ is the data to be observed when the model parameters are $\theta$, and $d$ is the experimental design to be applied.

The utility function $u(d,x,\theta)$ is of particular interest here because it describes the ``worth'' (based on the experimental aims) of choosing the proposed experimental design $d$ yielding data $y$ from a model with parameters $\theta$ \citep{Ryan2016}. This ``worth'' is usually focused on information gain rather than management outcomes, but the mathematical formulae used in optimal design do not require this to be the case. We will here make use of the broad range of what types of ``worth'' the utility function could capture, to establish the mathematical link between VoI and optimal design.

The \textit{expected} utility $u(d)$ of applying design $d$ can be calculated from $u(d,x,\theta)$. This calculation typically requires an expectation that is taken over both the prior distribution $p(\theta)$ and the future data $x$. It is given by 
\begin{equation} \label{eq:utility}
  u(d) =  \E_{\,\theta,x} [u(d,x,\theta)] = \displaystyle\int_x \displaystyle\int_\theta \, u(d,x,\theta) \, p(x|d,\theta) \, p(\theta) \, \mathrm{d}\theta \, \mathrm{d}x,
\end{equation}
where $p(x|d,\theta)$ is the likelihood function of the observed data given that the design $d$ is applied. The design $d^*$ that maximises the expected utility function $u(d)$ over all proposed experimental designs $d$ is given by:
\begin{equation}
d^*  = \underset{d}{\argmax} \;  u(d). \label{eq:design}
\end{equation}
Thus, the optimal design $d^*$ would be recommended as the experimental design that should be selected.

If, however, the utility function already accounts for all possible states $\theta$ of the system (see \citealt{Drovandi2013}, for example), then this utility function can be written as $u(d,x)$, and Equation~\eqref{eq:utility} simplifies to 
\begin{equation} \label{eq:utilityVOI}
u(d) =  \E_{x} [u(d,x)]  =  \displaystyle\int_x \, u(d,x) \, p(x|d) \, \mathrm{d} x.
\end{equation}
Recall that, in Section \ref{sec:VSIx}, we found that the traditional VoI metric for an imperfect measurement, EVSI, is equal to the expectation over all measurement outcomes $x$ of the VSI, a new metric that we introduce in the present work. Comparison of this finding to Equations \eqref{eq:design} and \eqref{eq:utilityVOI} reveals that traditional VoI calculations are a specific case of Bayesian optimal design, where (1)~there is one design $d$ representing the proposed measurement being considered in VoI analysis; (2)~the model parameters $\theta$ in optimal design are equivalent to the system state $s$ in VoI analysis; and (3)~the utility function $u(d,x)$ is chosen to be equal to the expected VSI given a particular measurement outcome \(x\): 
\begin{equation} \label{eq:UtilityToVSI}
u(d,x) = \text{VSI}_x.
\end{equation}
Hence, there is a clear mathematical link between VoI and optimal design that is here quantifiable after defining VSI (Equation~\eqref{eq:VSIx}). (Note that we could have equally chosen $u(d,x) = \Delta \text{EV}_x$, since both $\E_x \{ \text{VSI}_x \}$ and $\E_x \{ \Delta \text{EV}_x \}$ yield EVSI, but the former is more interpretable in the context of decision-making.)

The link between VoI and optimal design described here also permits interesting extensions for common VoI calculations. For example, consider the situation in which there are multiple different experiments being proposed that each yield imperfect measurements. If these experiments are compared individually for their effects on decision-making, choosing the experiment with the highest EVSI is equivalent in optimal design to assigning each proposed experiment a different design $d$ in Equation~\eqref{eq:design}, and thereafter choosing the utility function $u(d,x)$ to be equal to $\text{VSI}_x(d)$ calculated for the associated experimental design $d$ and measurement outcome $x$. The second case study in the present paper provides an example of this extension (Section \ref{sec:Case2}).

\section{Case studies} \label{sec:CaseStudies}

\noindent Here we use two case studies to demonstrate the potential utility of the proposed new VoI metrics. The first case study is simple and informs management actions to protect a threatened frog population from disease, whilst the second case study is more complicated and informs management actions to increase the survival rate of turtles released from captivity. Both case studies were analysed in \cite{Canessa2015} for the traditional VoI metrics of EVPI and EVSI. We restrict our attention to the realistic case of imperfect measurements, so focus on comparing our new metrics to the corresponding EVSI values. For each of the case studies (Sections \ref{sec:Case1} and \ref{sec:Case2}), we describe the environmental management problem being addressed, summarise the traditional VoI metric findings, and thereafter explain how our four new metrics ($\Delta$EV$_x$, VSI$_x$, \(\sigma \text{VSI}\), and \(\text{rVSI}_\delta\)) complement EVSI to provide a comprehensive overview of the potential value of a measurement being considered by a decision-maker to inform their actions.

\subsection{Case study 1: Protecting frogs from disease outbreaks} \label{sec:Case1}

\subsubsection{Problem description}

\noindent The first case study concerns the risk of disease outbreak in the management of a threatened frog species  \citep{Canessa2015}. The frog species has a total population of 100 individuals, all of which are located at a disease-free site within a protected area. The decision-maker's goal is to maximise the total frog population within the protected area. One potential action to achieve this goal is to establish a `new' population of the frog species by translocating some individuals to a new unoccupied site that is also within the protected area. The frog species management needs to be considered in light of two possible system states: the disease is present at the new site (state~$s_1$) or absent at the new site (state~$s_2$). The decision-maker currently believes there is a 50\% chance that the disease is present at the new site, i.e.\ $p(s_1) = p(s_2) = 0.5$. For simplicity, there are two possible actions being considered: translocate 50 of the 100 frogs to the new site (action~$a_1$), or do nothing (action~$a_2$). If nothing is done, the total frog population will remain as 100 individuals because the disease is not present at the site where the frog population currently resides. If the frogs are translocated and the new site is disease-free, the total population is expected to increase to 135 frogs. If the frogs are translocated and the disease is present at the new site, the total population is expected to reduce to 55 frogs.

Before choosing whether to translocate the frogs or not, there is an imperfect measurement that can be performed to test for the presence or absence of disease at the new site. The mean sensitivity (probability of the test being positive when disease is present, i.e.\ true positive) of the test is 0.73, whilst the specificity (probability of the test being negative when disease is absent, i.e.\ true negative) is 0.94. The decision-maker wishes to know whether to perform the test or not prior to choosing their management action (translocate frogs or do nothing).

\subsubsection{Outcomes from the traditional VoI metrics}

\noindent We here briefly summarise the relevant outcomes from traditional VoI metrics for this frog conservation case study (see Section S2.1 for further mathematical details). First, doing nothing (action $a_2$) is the best action if the decision-maker is risk-neutral and does not do the test; this yields an expected total frog population of 100 individuals (according to the metric EV$_{\text{uncertainty}}$). Second, if the test is performed, the expected frog population is 110.4 individuals (EV$_{\text{less uncertainty}}$), so the expected value of the test is 10.4 individuals (EVSI). An expected gain in 10.4 frogs might convince the decision-maker to do the test before choosing a management action.

\subsubsection{Outcomes from the new VoI metrics}
\noindent Even for this simple case study, the new VoI metrics we propose yield relevant insights (Table \ref{table:CS1results}). Whilst the EVSI is 10.4 frogs, the uncertainty in the value of the test is 8.4 frogs (according to the new metric $\sigma \text{VSI}$), thus yielding a projected value for the test of 10.4$\pm$8.4 frogs. Because the uncertainty in the value of test is quite high relative to the expected value of the test, this uncertainty might make a decision-maker more hesitant to do the test.

\renewcommand{\arraystretch}{1.3}
\begin{table}[!ht]
\centering \scriptsize
\caption{Summary of calculated values for traditional and new VoI metrics for the first case study of protecting frogs from disease outbreaks (see Section \ref{sec:Case1} for case study details, and Section S2.2 for mathematical details). Full mathematical definitions of all metrics are provided in Sections \ref{sec:def} and \ref{sec:VOIanalysis}. Notice that the 39.5\% risk of the proposed measurement yielding no value (VSI$_x=0$) is hidden if only the traditional VoI metric of EVSI is reported. *The values of metrics EVSI and $\sigma$VSI are independent of test outcome.\\}
\label{table:CS1results}
\begin{tabular}{p{0.5\linewidth} ccc}
\hline
\multirow{2}{*}{\textbf{Metric description}} & \textbf{Metric} & \textbf{Test positive} & \textbf{Test negative}  \\ 
& \textbf{symbol}  & \textbf{for disease} ($x_1$) & \textbf{for disease} ($x_2$)  \\ \hline
Shift in expected value of the system & $\Delta \text{EV}_x$ & 0 & 17.2  \\ 
Value of sample information & $\text{VSI}_x$ & 0 & 17.2  \\ 
Probability of observing the measurement outcome & $p(x)$ & 0.395 & 0.605  \\  
Expected value of sample information & $\text{EVSI}$ & 10.4* & 10.4*  \\ 
Standard deviation of the value of sample information & $\sigma \text{VSI}$ & 8.4* & 8.4* \\ \hline
\end{tabular}
\end{table}
\renewcommand{\arraystretch}{1}

If the test is negative for disease at the new site, which has a 60.5\% chance of occurring, the management action \textit{will} change -- from do nothing (action $a_2$) to translocation (action $a_1$)\,-- and the measurement has an expected value of 17.2 frogs (VSI$_x = 17.2$). In this case, the expected frog population after translocation will be 117.2 individuals, so there is an increase in the expected frog population after measurement ($\Delta \text{EV}_x = 17.2$). On the other hand, if the test is positive for disease at the new site, which has a 39.5\% chance of occurring, the management action will not change -- no frogs will be translocated (action $a_2$) -- so the measurement has no value (VSI$_x = 0$). In addition, the frog population will remain as 100 individuals, so there is no expected change in the total frog population ($\Delta \text{EV}_x = 0$). If this outcome is described in the context of risk aversion, there is a 39.5\% chance that the measurement yields no value (rVSI$_0$ = 0.395). This finding could be worrying for a risk-averse decision-maker because it is a reasonably high probability of a low-value measurement outcome. However, this finding is hidden when reporting only the traditional VoI metrics.

This simple example demonstrates that there is utility in reporting the new VoI metrics proposed here alongside traditional VoI metrics when a decision-maker is choosing whether to perform an additional measurement (or not) prior to action. We also note that, in this case study, $\Delta \text{EV}_x = \text{VSI}_x$ for all measurement outcomes $x$ (Table \ref{table:CS1results}); this possibility was noted in Section \ref{sec:VSIx}. In the second, more complicated, case study we will show that this is not true in general.

\subsection{Case study 2: Improving survival of turtles released from captivity} \label{sec:Case2}
\subsubsection{Problem description}

\noindent The second case study concerns a species conservation project that wishes to reintroduce a population of captive-bred turtles into the wild \citep{Canessa2015}. The recommended age range of released turtles is between 3 and 5 years old inclusive. Prior knowledge suggests that post-release survival of turtles may be age-dependent; there are three pre-existing hypotheses for how the survival of turtles depends on their age at time of release from captivity. Briefly, these three hypotheses are that there is no negative post-release effect of reintroduction on turtle survival (state $s_1$), the post-release effect decreases with turtle age (state $s_2$), or the post-release effect increases with turtle age (state $s_3$). The three potential management actions that the decision-maker is considering to implement are to release a very large number of captive turtles when they are three years old (action $a_1$), four years old (action $a_2$) or five years old (action $a_3$). The decision-maker wishes to maximise the survival probabilities of wild turtles at six years of age, as a proxy for long-term survival of wild turtles. How these probabilities depend on the system state $s$ and chosen management action $a$ are described in Table S3.1.

Prior to choosing the management action, the decision-maker is considering whether to perform one of three potential experiments to improve knowledge of the age-dependent survival of released turtles. All three different experiments being considered are trial releases of ten turtles into the wild for one year. This choice of experiments differs from that assumed in \cite{Canessa2015}, so our results differ from that work but are correct for our assumptions (S.\ Canessa, personal communications). All our computed outcomes, for the traditional and new VoI metrics for this turtle reintroduction case study, were obtained using the MATLAB code in Supplementary Material File S1.

For our choice of experiments, the difference between these experiments is the age of turtles being released: 3-year old turtles (experimental design $d_1$), 4-year old turtles (design $d_2$) or 5-year old turtles (design $d_3$). The survival probabilities of a single turtle in the wild for one year, $p_{\text{survival}}(d,s)$, depends on both the turtle's age at time of release (design $d \in \{ d_1,d_2,d_3 \}$) and which of the three different hypotheses $s \in \{ s_1,s_2,s_3 \}$ about the post-release survival rate of turtles is true (Table S3.2). For each trial release, there are eleven potential measurement outcomes -- between zero and ten turtles inclusive will survive at the end of one year (i.e.\ $x \in \{0,1,2,...,9,10 \}$). A binomial probability distribution can be used on a single turtle's survival rate (Table S3.2) to calculate the probability of each of the eleven potential measurement outcomes $x$ for each of the three proposed experiments.

\subsubsection{Outcomes from the traditional VoI metrics}
\noindent We here briefly summarise the relevant outcomes from traditional VoI metrics for this turtle reintroduction case study. First, if the decision-maker is risk-neutral and does not perform any trial releases prior to choosing a management action, releasing a large number of 4-year old turtles is the best action. This will yield an expected survival probability for turtles at 6 years of age of 62.0\% (according to the metric EV$_{\text{uncertainty}}$). Second, doing any of the three proposed trial releases will improve the expected survival probability for turtles at 6 years of age, by differing amounts. If the trial release is performed with 3-, 4- or 5-year old turtles, the expected survival probability for turtles at 6 years of age increases to 62.4\%, 64.1\% or 65.4\% respectively (EV$_{\text{less uncertainty}}$ calculated for each experimental design). Thus, the value of each trial release is an increase in expected survival probability for 6-year old turtles of 0.4\%, 2.1\% and 3.4\% respectively (EVSI calculated for each experimental design). The best experiment based on maximum EVSI is therefore a trial release of ten 5-year old turtles.

Note that the choice of which trial to perform, based on which trial has the largest EVSI, is mathematically identical to choosing the design with maximum utility if this were treated as an optimal design problem (Equation \eqref{eq:design}); thus this case study highlights the mathematical link between VoI and optimal design described in Section \ref{sec:optdesign}.

\subsubsection{Outcomes from the new VoI metrics}

\begin{figure}[p]
\centering
\includegraphics[width=\textwidth]{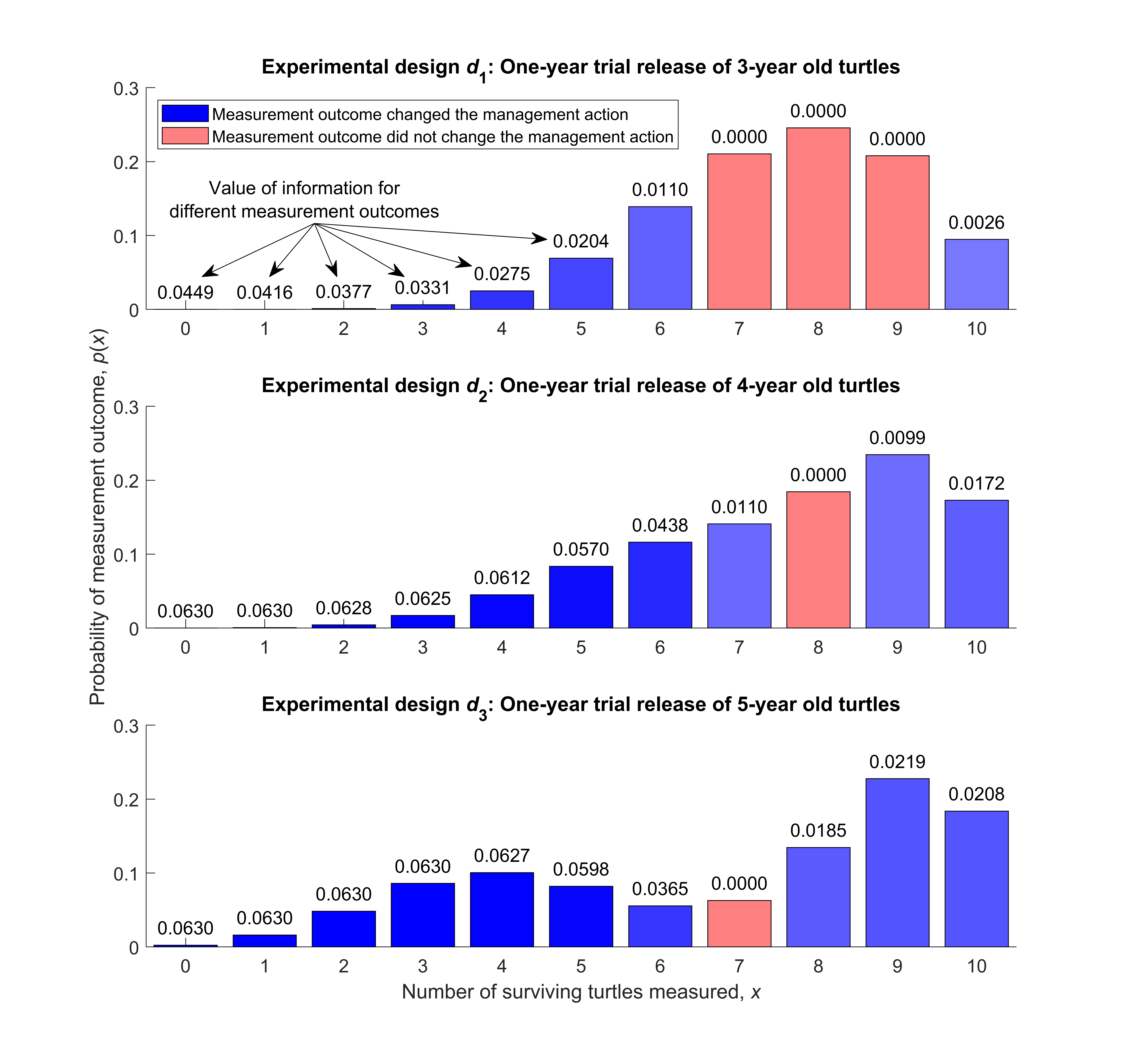}
\caption{Probability of measurement outcomes $p(x)$ ($y$-axis) and value of sample information VSI$_x$ (numbers above each bar) for all possible measurement outcomes $x$ of the one-year trial release of 3- (top), 4- (middle) and 5-year old turtles (bottom), for the second case study of improving survival of turtles released from captivity (see Section \ref{sec:Case2} for case study details). Bars in blue indicate that the measurement outcome $x$ yields a positive value of information (VSI$_x>0$) so the management action will change; the darker the shade of blue, the greater the value of the measurement outcome for the action. Bars in light red indicate that the measurement outcome $x$ yields no value (VSI$_x=0$) so the management action will not change. In this figure the symbol $d_i$, where $i=1,2,3$, refers to the $i$th experimental design or strategy $d$, and it is assumed that the best one design out of $d_1$, $d_2$ or $d_3$ may be implemented. Notice that the probability of zero information value across all potential measurement outcomes, $\text{rVSI}_0 = p(\text{VSI}_x=0)$ (sum of heights of all light red bars), is smallest for the one-year trial of releasing 5-year old turtles (design $d_3$, bottom plot).}
\label{fig:CaseStudy2VoIfigure}
\end{figure}

\noindent New metrics we introduce that are independent of measurement outcome ($\sigma$VSI and rVSI$_\delta$) complement the conclusion based on maximum EVSI that the trial release consisting of ten 5-year old turtles in the best choice of the three proposed trials. First, the new metric $\sigma$VSI demonstrates that the one-year trial of releasing 5-year old turtles is the only experiment of the three that does not have an uncertainty in its value for decisions ($\sigma$VSI) that is higher than its expected value for decisions (EVSI; see Table \ref{table:CaseStudy2VoISummary}). This indicates that the range of potential value gains from the one-year trial of releasing 5-year old turtles, is mostly constrained to positive values. As a result, the decision-maker can be confident that the outcome of this trial will likely be valuable to inform their management action selection. Second, the risk of zero information value, evaluated using the risk metric rVSI$_\delta$ for a threshold VSI of $\delta = 0$, rVSI$_0$, is smaller by a factor of 3-10 fold for the one-year trial of releasing 5-year old turtles compared to the other proposed experimental designs  (Table \ref{table:CaseStudy2VoISummary} and Figure \ref{fig:CaseStudy2VoIfigure}). This indicates that the one-year trial of releasing 5-year old turtles is the best experiment to choose if the decision-maker is risk-averse to zero-value measurement outcomes.

\renewcommand{\arraystretch}{1.3}
\begin{table}[ht!]
\centering
\caption{Calculated values for traditional and new VoI metrics, that are independent of measurement outcome, for the second case study of improving survival of turtles released from captivity (see Section 3.2 for case study details). In this table the symbol $d_i$, where $i=1,2,3$, refers to the $i$th experimental design or strategy $d$, and it is assumed that the best one design out of $d_1$, $d_2$ or $d_3$ may be implemented. Full mathematical definitions of all metrics are provided in Sections \ref{sec:def} and \ref{sec:VOIanalysis}. Notice that the metrics EVSI, $\sigma$EVSI and rVSI$_0$ together suggest that the decision-maker should choose to do the one-year trial of releasing 5-year old turtles (design $d_3$).  \\}
\label{table:CaseStudy2VoISummary}
\begin{tabular}{ccc}
\hline
\multirow{2}{*}{\textbf{Experimental design} $d$} & \multicolumn{2}{c}{\textbf{VoI metrics}} \\ \cline{2-3}
 & EVSI $\pm$ $\sigma$VSI & rVSI$_0$ \\ \hline
One-year trial release of 3-year old turtles ($d_1$) & 0.0042 $\pm$ 0.0179 & 66.4\% \\
One-year trial release of 4-year old turtles ($d_2$) & 0.0209 $\pm$ 0.0259 & 18.4\% \\
One-year trial release of 5-year old turtles ($d_3$) & 0.0342 $\pm$ 0.0239 & 6.3\% \\ \hline
\end{tabular}
\end{table}
\renewcommand{\arraystretch}{1}

If the decision-maker is unsure whether they should perform the one-year trial of releasing 5-year olds or not perform any trial at all, the new metrics we introduce are useful here as well. The obtained value of rVSI$_0$ = 6.3\% (Table \ref{table:CaseStudy2VoISummary}) indicates that the one-year trial of releasing 5-year old turtles has a 93.7\% chance of changing the management action chosen by the decision-maker. In other words, this one-year trial has a 93.7\% chance of improving the design of the large-scale turtle reintroduction. Because this means that the trial has a 93.7\% chance of increasing the expected long-term survival of wild turtles, the additional metric of rVSI$_0$ here may be a critical piece of information that convinces the decision-maker of whether to perform the trial or not.

Next, using the measurement outcome-dependent metric VSI$_x$, we find that only one outcome of the trial of releasing 5-year old turtles -- measurement of seven surviving turtles~-- yields zero information gain for the purpose of improving decisions (light red bar in bottom panel of Figure \ref{fig:CaseStudy2VoIfigure}). Thus, this measurement outcome of the trial, which has a probability of 6.3\% (Table \ref{table:CaseStudy2FullResults5YearOlds}), is \textit{solely} responsible for the non-zero value of the risk metric rVSI$_0$ for the trial of releasing 5-year old turtles. Similarly, only one outcome of the trial of releasing 4-year old turtles -- measurement of eight surviving turtles -- yields zero information gain (light red bar in middle panel of Figure \ref{fig:CaseStudy2VoIfigure}), although this outcome has a higher probability of 18.4\% (Tables \ref{table:CaseStudy2VoISummary} and S3.6) compared to the equivalent zero-value measurement outcome of the trial of releasing 5-year old turtles. On the other hand, the trial of releasing 3-year old turtles will yield zero information gain for the purpose of improving decisions if the outcome of the trial is 7, 8 or 9 surviving turtles (light red bars in top panel of Figure \ref{fig:CaseStudy2VoIfigure}); these outcomes possess probabilities of 21.1\%, 24.6\% and 20.8\% (Table S3.4), together yielding a risk of zero information gain of 66.4\% (Table \ref{table:CaseStudy2VoISummary}). Hence, the metric VSI$_x$ explains how the risk of low-value measurement outcomes is \textit{spread} across the different potential measurement outcomes of a proposed experiment.

\renewcommand{\arraystretch}{1.3}
\begin{table}[ht!]\footnotesize
\centering
\caption{Calculated values for the new VoI metrics that are dependent on measurement outcomes, alongside the probabilities of these outcomes and corresponding flow-on effects to the recommended management action, for the third proposed experimental design ($d_3$: release 5-year old turtles for one year) in the second case study of improving survival of turtles released from captivity (see Section \ref{sec:Case2} for case study details). Negative values of $\Delta \text{EV}_x$ indicate that the measurement outcome will cause us to downgrade our estimate of the expected value of the system (i.e.\ this means that, prior to measurement, the expected value of the system was being overestimated). Positive values of VSI$_x$ indicate that the new measurement has helped to improve the action that is chosen for this system. Equivalent calculated values for the first and second proposed experimental designs of the second case study are provided in Tables S3.4 and S3.6 respectively. *Note that this column is indicating the recommended change in the design of the \textit{large-scale} turtle reintroduction (management action) in response to the outcome of the \textit{one-year trial} of releasing 5 year-old turtles (measurement). \\}
\label{table:CaseStudy2FullResults5YearOlds}
\begin{tabular}{ccccc} 
\hline
\multicolumn{5}{c}{ \textbf{Experimental design $d_3$: One-year trial release of 5-year old turtles}} \\ \hline
\textbf{Number of} & \textbf{Shift in expected} & \textbf{Value of sample} & \textbf{Probability of}  & \textbf{Does the recommended}  \\ 
\textbf{surviving} & \textbf{value of the} & \textbf{information,} & \textbf{measurement} & \textbf{action change after the} \\ 
\textbf{turtles,} $x$ & \textbf{system,} $\Delta \text{EV}_x$ & $\text{VSI}_x$ & \textbf{outcome,} $p(x)$ & \textbf{measurement outcome?*} \\
\hline
        0 & --0.0730 & 0.0630 & 0.0024 & Yes, release 3-year olds instead \\
        1 & --0.0730 & 0.0630 & 0.0161 & Yes, release 3-year olds instead \\
        2 & --0.0730 & 0.0630 & 0.0484 & Yes, release 3-year olds instead  \\
        3 & --0.0730 & 0.0630 & 0.0860 & Yes, release 3-year olds instead  \\
        4 & --0.0729 & 0.0627 & 0.1006 & Yes, release 3-year olds instead  \\
        5 & --0.0714 & 0.0598 & 0.0821 & Yes, release 3-year olds instead  \\
        6 & --0.0575 & 0.0365 & 0.0556 & Yes, release 3-year olds instead \\
        7 &   0.0227 & 0      & 0.0629 & No, still release 4-year olds  \\
        8 &   0.0982 & 0.0185 & 0.1345 & Yes, release 5-year olds instead  \\
        9 &   0.1128 & 0.0219 & 0.2276 & Yes, release 5-year olds instead  \\
        10 &  0.1165 & 0.0208 & 0.1838 & Yes, release 5-year olds instead \\ \hline
\end{tabular}
\end{table}
\renewcommand{\arraystretch}{1}

For a decision-maker wanting to avoid low but positive measurement outcomes, the risk of low value of sample information, rVSI$_\delta$, may need to be calculated for values of $\delta$ larger than zero. Calculating such risk metrics can be illustrated using Figure \ref{fig:CaseStudy2VoIfigure}. Blue and red bars in Figure \ref{fig:CaseStudy2VoIfigure} indicate measurement outcomes $x$ that yield positive (VSI$_x > 0$) and zero (VSI$_x = 0$) values of information, respectively, with the value of VSI$_x$ stated above each bar. For example, if the decision-maker wants the trial's value to be an expected increase in long-term survival rate of wild turtles of more than 5\%, the probabilities $p(x)$ (\textit{y}-axis) associated with measurement outcomes $x$ (\textit{x}-axis) that yield VSI values that are less than or equal to 0.05 can be added together to calculate the risk that this goal is not met. For the one-year trial release of 5-year old turtles (Figure \ref{fig:CaseStudy2VoIfigure}, bottom panel), VSI$_x < 0.05$ if 6, 7, 8, 9 or 10 turtles survive the trial. These five outcomes possess probabilities of 0.0556, 0.0629, 0.1345, 0.2276 and 0.1838 respectively (Figure \ref{fig:CaseStudy2VoIfigure} and Table \ref{table:CaseStudy2FullResults5YearOlds}); which together add to 0.66. Thus, the probability of the trial yielding an expected increase in survival rate of released turtles that is less than 5\%, is 66\% (mathematically, rVSI$_{0.05} = 0.66$). Based on this information, a decision-maker that is risk-averse to such low-value measurement outcomes might reconsider whether or not they should perform the one-year trial release of 5-year old turtles.

Finally, this case study provides an example of when the metrics $\Delta \text{EV}_x$ and $\text{VSI}_x$ are not equal to each other (Table \ref{table:CaseStudy2FullResults5YearOlds}). Such a situation can arise, for example, when new information gained from an experiment causes the estimates of the highest achievable expected value of the system to be downgraded (i.e.\ negative $\Delta \text{EV}_x$) whilst simultaneously causing a better management action to be chosen (i.e.\ positive VSI$_x$).

For the trial release of 5-year old turtles, measurement outcomes of less than seven out of ten turtles surviving in the wild for one year, possess negative $\Delta \text{EV}_x$ and positive VSI$_x$ (Table \ref{table:CaseStudy2FullResults5YearOlds}). When these measurement outcomes are obtained, the hypothesis of reduced post-release survival of older turtles is given greater weight (i.e.\ system state $s_3$, see associated upward-pointing arrows in Table S3.3) compared to the other two hypotheses about the system. The expected long-term survival rate of wild turtles is therefore downgraded (negative $\Delta$EV$_x$) because the now more strongly weighted hypothesis about the system ($s_3$) is, overall, more pessimistic about long-term wild turtle survival rates than the other two hypotheses (Table S3.1). However, a potentially better management action will be chosen with this improved knowledge gained from the trial. According to the hypothesis of reduced post-release survival of older turtles (system state $s_3$), the long-term survival of wild turtles is greatest if they are released from captivity at the age of 3 years old (Table S3.1), so the recommended management action changes from releasing 4-year old turtles to instead releasing 3-year old turtles (positive VSI$_x$). Using similar logical deductions, the obtained values of $\Delta \text{EV}_x \neq \text{VSI}_x$ can be explained for all measurement outcomes of all three proposed one-year trial releases (i.e.\ of 3-, 4- and 5-year old turtles) using information presented in Tables \ref{table:CaseStudy2FullResults5YearOlds}, S3.1 and S3.3-S3.7. This information only needs to be communicated if it is necessary to \textit{explain why} it is possible for the value of a measurement to be positive and the system's expected value to reduce at the same time.

\section{Discussion} \label{sec:discussion}

\subsection{New and traditional VoI metrics together yield a comprehensive view of measurement value}
\label{sec:new_vs_traditional}
\noindent Environmental decision-making, whether it is about choosing to monitor and/or to act, always carries some element of risk \citep{Bennett2018}; this is especially true when the consequences of an incorrect decision are ecologically catastrophic \citep{Martin2012}. Gathering new information can therefore be a risky activity \citep{Freixas1984}, and the relationship between risk aversion and the desire of a decision-maker to measure before choosing an action can be unavoidably complicated \citep{Bleichrodt2020}. The present work does not intend to resolve the complexities of the human element of decision-making; rather we show how additional metrics which complement the traditional metric of VoI for an imperfect measurement (EVSI) can explicitly describe the risk and uncertainty of such a proposed measurement. The new metrics can also complement the traditional metric of VoI for a perfect measurement, EVPI, via a mathematical simplification described at the start of Section \ref{sec:VOIanalysis}. Hence the approach to VoI analysis presented in this paper yields a comprehensive set of information for the decision-maker to determine the next step forward to best manage their system, regardless of both the monitoring activities that are available and the cognitive processes that are invoked in the decision-making process.

First, the metric $\sigma$VSI describes the uncertainty in the value of information gained from a measurement, so provides an excellent complement to EVSI. We suggest that these two metrics (traditional and new) can be reported together as EVSI$\pm\sigma$VSI (Table \ref{table:CaseStudy2VoISummary}). Second, the metric rVSI$_\delta$ expresses the risk that the proposed measurement will yield an information value that is viewed as not sufficiently useful for decisions (quantified as VSI less than or equal to a threshold value $\delta$). In the specific case of \mbox{$\delta=0$} this risk metric can also be interpreted as the risk that the information gained from the proposed measurement does not change the decision that would otherwise have been made if the measurement was not performed. Hence, in absence of additional knowledge about the decision-maker's preferences, $\delta = 0$ is an obvious default choice; however, if the decision-maker's tolerance to low-value measurement outcomes is known, a specific value of $\delta \geq 0$ can instead be used. Finally, the metrics $\Delta \text{EV}_x$ and VSI$_x$, which are both dependent on potential outcomes $x$ of the proposed measurement, disentangle how the proposed measurement may improve the estimates of the maximum achievable expected value of the system, and improve decision-making about this system, respectively.

Whilst the traditional metrics of VoI are expectations over measurement outcomes and therefore inherently assume risk-neutrality of the decision-maker with regards to measurement outcomes \citep{Yokota2004}, our additional metrics can account for the risk tolerance of the decision-maker to low-value measurement outcomes. Acknowledging such risks by explicitly calculating their probability of occurrence via the metric rVSI$_\delta$, may be sufficient to help a risk-averse decision-maker to choose whether to proceed with a measurement or not. For example, in the first case study presented here, the traditional metric EVSI indicated that the proposed measurement (a test for disease at a potential frog translocation site) will yield an average saving of approximately 10 additional frogs, whilst the new metric rVSI$_0$ highlighted that there is a 40\% chance that the measurement is valueless (Section \ref{sec:Case1}). On the other hand, in the second case study, the traditional metric EVSI indicated that the most valuable proposed measurement (a one-year trial release of ten 5-year old turtles) will yield an increase in the expected long-term survival probability for wild turtles of 3\%, whilst rVSI$_0$ indicates that there is only a 6\% chance that the measurement is valueless (Table \ref{table:CaseStudy2FullResults5YearOlds}). In both case studies, the complementary metrics (traditional and new) provide different and potentially pivotal perspectives on the measurement's value to the decision-maker. In any case, it should be noted that VoI analysis by itself cannot influence potential measurement outcomes; however, the probability of specific measurement outcomes can be forecasted (using e.g.\ Bayes' theorem and conditional probability formulae), and VoI analysis can be subsequently used to forecast the flow-on effects of these potential outcomes on decision-making (Table \ref{table:CaseStudy2FullResults5YearOlds}).

Overall, although the traditional and new VoI metrics provide different perspectives on measurement value, there are advantages and disadvantages of these metrics when presenting them to a decision-maker. Traditional VoI metrics such as EVPI and EVSI are expectations so are ``average values'' which can be understood clearly. If a proposed measurement yields imperfect information, reporting of both EVPI and EVSI may be useful to indicate how ``imperfect'' the measurement is for the purposes of decision-making. However, we argue in the present work that these ``average'' metrics hide the uncertainty of how information value depends on measurement outcome, which could disappoint a decision-maker if a low-value outcome is indeed obtained. The new VoI metric of $\sigma$VSI addresses this issue, as it summarises the uncertainty of information value in one number. The new VoI metric VSI$_\delta$, when set to the default choice $\delta=0$ (or user-defined choice $\delta>0$), is a convenient metric to explicitly capture the risk of a measurement yielding no (or low) value for decision-making, so will likely be of interest as an additional metric when risk of low-value measurement outcome is a concern to the decision-maker. The other two new VoI metrics of $\Delta$EV$_x$ and VSI$_x$, because of their dependence on measurement outcome $x$, will likely only be beneficial when the decision-maker specifically seeks comprehensiveness of the VoI analysis. Hence, if the proposed measurement is not guaranteed to yield perfect information, it may be that EVPI, EVSI, $\sigma$VSI and rVSI$_0$ are metrics that are convenient to report together, because they summarise substantial information about the potential information for decision-making gained from a measurement, in only four numbers.

\subsection{The new VoI metrics are informative for addressing other risks of measurement}
\label{sec:discussion_other_risks}

\noindent VoI analysis, when including the new metrics discussed here, can address other potential concerns of the decision-maker beyond the risk of low-value measurement outcomes, as follows. First, if the decision-maker is considering to perform the measurement only because they desire more evidence that their currently proposed action (i.e.\ best action under uncertainty) is the best, the metric rVSI$_0$ can instead be interpreted as the probability that the measurement will confirm that the decision-maker is choosing the best action. This is a novel application of rVSI$_0$ that may be useful for a cautious decision-maker. Second, if the decision-maker is risk-averse to undesirable system outcomes, then the value function $V(a,s)$ used in VoI analysis can be scaled by an appropriately-chosen monotonic function of the desired system value \citep{Hilton1981}. In this second case, the new metrics can still be used, thus extending VoI analysis to address risk aversion to low-value measurement outcomes and low-value system outcomes simultaneously. However, care would need to be taken when calculating and reporting values of both the traditional (EVSI) and new metrics ($\sigma$VSI, rVSI$_\delta$, VSI$_x$ and $\Delta \text{EV}_x$) to ensure that their outcomes are not misinterpreted due to the rescaling of $V(a,s)$ required for addressing risk aversion to low-value system outcomes.

In the case studies discussed in the present work, VoI analysis was considered for \textit{static} problems in which there is one set of management actions to choose from at a particular time, and potential data collection activities that could inform the choice of actions at that time point only (Figure \ref{fig:timeline}). In this context, static VoI analysis is particularly helpful as a fast prototyping tool when measurements are identified as valueless, since actions can then be chosen without collecting additional information. If, in contrast, the VoI analysis identifies measurements as having sufficient potential value, static VoI analysis may itself be insufficient for decision-making if there is a long term management goal for the system of interest. This issue has motivated VoI analysis approaches in a more \textit{dynamic} setting, by analysing a sequence of (potential future) management actions over time that is affected by corresponding (potential future) changes in the environmental system, altogether as one large VoI problem \citep{Xiao2019}. The latter approach is able to yield a comparison of different adaptive management strategies \citep{Chades2017}. We acknowledge that the new metrics we introduce here were designed in the context of static VoI problems; whether these metrics are directly transferable to dynamic VoI problems may be worthy of consideration in future research.

\subsection{Consequences of the mathematical link between VoI analysis and optimal design}
\noindent Both VoI and optimal design are decision-theoretic approaches for estimating the expected benefits of collecting further information about a system \citep{Jackson2022}. The metric VSI$_x$ introduced in the present work permits a clear mathematical link to be made between VoI and optimal design (Section \ref{sec:optdesign}). Specifically, VoI is a special case of optimal design in which the utility function being maximised is the value of sample information associated with the proposed measurement (Equation \eqref{eq:UtilityToVSI}). This means that techniques developed in the field of optimal design \citep{Ryan2016} may find new applications in VoI analysis; these connections remain unexplored but potentially ripe for future work.

Whilst VoI specifies the value of a proposed measurement for the improvement in decision-making, optimal design may not always be used for this specific purpose. For example, a common utility function chosen in optimal design is based on the precision in knowledge of the system after the measurement (e.g.\ \citealt{Drovandi2013}). In this example, optimal design specifies the value of the proposed measurement for the improvement in knowledge of the system, which is a related but not identical goal to improvement in decision-making. These differences in objective have led some authors to argue that VoI and optimal design have unavoidably different goals \citep{McDonaldMadden2010}. 

There may indeed be a linguistic resolution to this issue. Notice that VoI analysis aims to quantify the value of a measurement for improved decision-making (via an uncommon utility function within the optimal design framework), but optimal design (with a commonly chosen utility function) aims to quantify the value of a measurement for information gain. Hence, it may be that, instead of treating this distinction as VoI versus optimal design, one should instead treat this distinction as VoI for improved decision-making versus VoI for improved knowledge. If this linguistic change is made, both approaches may be referred to as optimal design and VoI simultaneously -- with the chosen utility function determining whether the goal of the design aims specifically to improve decision-making or improve knowledge. This resolution is most suitable for the case of static problems (see Section \ref{sec:discussion_other_risks}) in which there is \textit{one} time at which the measurement may occur and \textit{one} (later) time at which the management action may occur (Figure \ref{fig:timeline}). For dynamic problems that require sequential decision-making, mathematical techniques for adaptive management are likely to be more appropriate \citep{Chades2017}.

For both case studies considered here, calculating VoI was computationally inexpensive. However, it is well-known in the optimal design literature that a potential challenge in applying optimal design of experiments (and therefore also in VoI analysis) is the computational feasibility for large and complex problems such as those representing real and complicated ecosystems. Because the expected utility function for a design must often be estimated using Monte Carlo integration methods, computationally intensive stochastic optimisation schemes, such as simulated annealing \citep{Kirkpatrick1983} or approximate coordinate exchange \citep{Overstall2017} are required. While improving the computational efficiency of these schemes is still an ongoing area of research, there are some common approaches to approximate utility functions including variational inference \citep{Bei2017}, integrated nested Laplace approximation \citep{Rue2009} and generalised estimating equations \citep{Woods2011,Gill2022}. Recent advances in variance reduction \citep{Jacob2020}, multilevel Monte Carlo \citep{Giles2015,Jasra2019,Warne2022} and deep reinforcement learning \citep{Blau2022} can dramatically improve the performance of complex Monte Carlo integrals; such methods have the potential to enable optimal design and VoI to scale to complex conservation projects.

\subsection{Good decisions, fast decisions}

\noindent Since decisions about environmental systems sometimes must be made rapidly \citep{Martin2012}, there may be both limitations to what information can be gathered to inform a VoI analysis, and time pressures on communication to decision-makers prior to an action being chosen. If the information gathering process is not sufficiently rapid, there is also the risk that a subsequent delay in management actions is costly to the system  being managed \citep{Grantham2009}. Hence, there may be time restrictions imposed on what measurements can be feasibly performed, and such measurements might therefore yield rapid but small information value compared to more comprehensive (and more time-consuming) monitoring campaigns. We below discuss two other considerations regarding whether/how a VoI analysis should be performed, and how the results of such analyses could be reported.

First, VoI analysis requires sufficient information to produce forecasts of system value in response to different management actions and measurement outcomes, and thus like other quantitative ecological forecasts \citep{Adams2020}, VoI analysis is data-hungry. The quantitative VoI analysis that is the subject of the present work (see Figure \ref{fig:timeline}) requires definition and quantification of system values $V(a,s)$ and their dependence on potential management actions $a$ and potential system states $s$, as well as quantification of prior probabilities $p(s)$ that the system is indeed in state $s$. When such information is available, VoI can be immediately and reliably undertaken, and if of interest can also prioritise which components of data-hungry forecasting models should be improved \citep{Strong2014}. When data are more scarce, qualitative approaches can instead be used \citep{ClarkWolf2022}; such approaches are currently being developed for VoI \citep{Rushing2020,Lawson2022,Stantial2023}. Such qualitative approaches place more emphasis on expert elicitation than on collecting quantitative data prior to gathering new information. It is somewhat of a ``chicken and egg'' problem that quantitative VoI requires quantitative data to have been previously collected to inform future quantitative data collection. Qualitative VoI, on other hand, makes use of expert elicitation to inform quantitative data collection, so overcomes this issue. Qualitative approaches for VoI should ideally be later replaced by quantitative approaches, once more data becomes available.

Second, the desire to simplify outputs of environmental analysis to one or a few indicators (e.g.\ \citealt{Wibeck2006,ElGibari2019}) may favour, in the context of VoI, strategic decisions about which metrics should be reported. This goal must be balanced by the increased transparency gained from commensurate reporting of uncertainty and risk \citep{Chess2005,Burgass2017} for environmental metrics that directly inform decision-making. What is considered ``best practice'' here will ultimately depend on the decision-maker's preferences. For example, as discussed in Section \ref{sec:new_vs_traditional}, one potential set of metrics to present is the four numbers of EVPI, EVSI, $\sigma$VSI and rVSI$_0$. These numbers together tell (1) a measure of how close to ``perfect'' the measurement is (comparing EVPI and EVSI), (2) estimated mean and uncertainty of VoI prior to performing the measurement (EVSI$\pm\sigma$VSI), and (3) the probability that the measurement will yield no change in the decision (rVSI$_0$). If reporting four VoI metrics (two traditional and two new) is considered to be too complicated to inform real-world decision-making, a subset of these metrics could be chosen, depending on what information is deemed to be of greatest importance. It may that only EVSI$\pm\sigma$VSI is reported (one traditional VoI metric and one new VoI metric) as this quantifies the estimated mean and uncertainty of the measurement's information value. On the other hand, if a more comprehensive VoI analysis is sought by the decision-maker, the metrics of EV$_x$ and VSI$_x$ may also be useful to report as they together indicate how the system's value \textit{and} the decision-maker's choice of action depends on measurement outcome.

\section{Conclusion}
\noindent In the present work we have introduced four new VoI metrics that can address the uncertainty and risk in information value associated with different outcomes of a proposed measurement activity. The new metrics can be used strategically to complement traditional VoI metrics when reporting to a decision-maker on the potential impact of a measurement activity on the improvement in subsequent management outcomes. Environmental decision-making is a challenging but increasingly critical endeavour; the more informed we are, the better these decisions will be.

\section*{Competing interests}
\noindent The authors declare that they have no competing interests.

\section*{Author's contributions}
\noindent MDA wrote the first draft of the manuscript, the code for numerical simulations, and analysed the simulation output. All authors contributed to review and editing of the manuscript.
 
\section*{Acknowledgements}
\noindent MDA, SAV and MPA acknowledge funding from the Australian Research Council (ARC) Discovery Early Career Researcher Award DE200100683. MDA, DJW and SAV also acknowledge funding from Queensland University of Technology's Centre for Data Science. MDA acknowledges funding from the ARC SRIEAS Grant SR200100005 Securing Antarctica's Environmental Future. KJH acknowledges funding from the ARC Discovery Early Career Researcher Award DE200101791. The authors thank Michael Bode, Stefano Canessa, Grace Heron, Kate Saunders and Ayesha Tulloch for fruitful discussions during the development of this manuscript.

\bibliographystyle{elsarticle-harv}
\scriptsize
\setlength\bibsep{0pt plus 0.3ex} 
\bibliography{Bibliography} 

\end{document}


\hfuzz=50pt
\hbadness=20000
\sloppy

\begin{frontmatter}
\title{Supplementary Material Text S1\protect\\ \textcolor{white}{.} \protect\\ Beyond expected values: Making environmental decisions using value of information analysis when measurement outcome matters}
\author[QUT,CDS,SAEF]{Morenikeji D. Akinlotan} \ead{morenikeji.akinlotan@qut.edu.au}
\author[QUT,CDS]{David J. Warne} \ead{david.warne@qut.edu.au}
\author[QUT,CDS,SAEF]{Kate J. Helmstedt} \ead{kate.helmstedt@qut.edu.au}
\author[QUT,CDS]{Sarah A. Vollert} \ead{sarah.vollert@connect.qut.edu.au}
\author[CSIRO_Iadine]{Iadine Chad\`{e}s} \ead{iadine.chades@csiro.au}
\author[QUT,CDS,USC]{Ryan F. Heneghan} \ead{rheneghan@usc.edu.au}
\author[Deakin,CSIRO_Hui]{Hui Xiao} \ead{hui.xiao@csiro.au}
\author[QUT,CDS,SAEF,SCC]{Matthew P. Adams\corref{cor1}} \ead{mp.adams@qut.edu.au}
\cortext[cor1]{Corresponding Author}
\address[QUT]{School of Mathematical Sciences, Queensland University of Technology, Brisbane QLD 4001, Australia}
\address[CDS]{Centre for Data Science, Queensland University of Technology, Brisbane QLD 4001, Australia}
\address[SAEF]{Securing Antarctica's Environmental Future, Queensland University of Technology, Brisbane QLD 4001, Australia}
\address[CSIRO_Iadine]{Land and Water Flagship, Commonwealth Scientific and Industrial Research Organisation, Dutton Park QLD 4102, Australia}
\address[USC]{School of Science, Technology and Engineering, University of the Sunshine Coast, Petrie QLD 4502, Australia}
\address[Deakin]{School of Life and Environmental Sciences, Deakin University, Burwood VIC 3125, Australia}
\address[CSIRO_Hui]{Commonwealth Scientific and Industrial Research Organisation, Queensland Biosciences Precinct, St Lucia QLD 4067, Australia}
\address[SCC]{School of Chemical Engineering, The University of Queensland, St Lucia QLD 4072, Australia}
\end{frontmatter}

\counterwithin{figure}{section}
\counterwithin{table}{section}
\counterwithin{equation}{section}

\renewcommand{\thesection}{S\arabic{section}}
\renewcommand{\theequation}{S\arabic{section}.\arabic{equation}}
\renewcommand{\thefigure}{S\arabic{section}.\arabic{figure}}
\renewcommand{\thetable}{S\arabic{section}.\arabic{table}}

\vspace{-1cm}

\clearpage

\section{Mathematical connections between EVSI, $\Delta$EV$_x$ and VSI$_x$}

\subsection{The expectation of $\Delta \mathrm{EV}_x$ over all measurement outcomes yields $\mathrm{EVSI}$}
\noindent In Section 2.2.1 of the main text, we stated that $\text{EVSI} = \E_x [ \Delta \text{EV}_x ]$. Here we show this derivation. From the definition of $\Delta \text{EV}_x$ in Equation (7), the expectation of $\Delta \text{EV}_x$ over all possible measurements $x$ is:
\begin{align}
  & \E_{x} \left[\Delta \text{EV}_{x} \right] \nonumber \\
= & \E_{x} \left[ \max_a \{ \E_{s|x} [V(a,s)] \}  - \max_a \{ \E_{s} [V(a,s)] \} \right] \nonumber \\
= & \E_{x} \left[ \max_a \{ \E_{s|x} [V(a,s)] \} \right]  - \E_{x} \left[ \max_a \{ \E_{s} [V(a,s)] \} \right] \nonumber \\
= & \E_{x} \left[ \max_a \{ \E_{s|x} [V(a,s)] \} \right]  - \max_a \{ \E_{s} [V(a,s)] \} \quad (\text{second term is independent of $x$})
\end{align}
which is the definition of EVSI in Equation (6).

\subsection{The expectation of $\mathrm{VSI}_x$ over all measurement outcomes yields $\mathrm{EVSI}$}
\noindent In Section 2.2.2 of the main text, we stated that $\text{EVSI} = \E_x [ \text{VSI}_x ]$. Here we show this derivation. From the definition of $\text{VSI}_x$ in Equation (9), the expectation of $\text{VSI}_x$ over all possible measurements $x$ is:
\begin{align}
& \E_{x} \left[\text{VSI}_{x} \right] \nonumber \\
= & \E_{x} \left[ \max_a \{ \E_{s|x} [V(a,s)] \}  - \E_{s|x} [V(a^*,s)] \right] \nonumber \\
= & \E_{x} \left[ \max_a \{ \E_{s|x} [V(a,s)] \} \right] - \E_{x} \left[ \E_{s|x} [V(a^*,s)] \right] \nonumber \\ 
= & \E_{x} \left[ \max_a \{ \E_{s|x} [V(a,s)] \} \right] - \E_{s} [V(a^*,s)] \; \qquad \quad \, (\text{by the law of total expectation})  \nonumber \\
= & \E_{x} \left[ \max_a \{ \E_{s|x} [V(a,s)] \} \right] - \max_a \{ \E_s [V(a,s)] \} \quad (\text{from $a^*$ definition, Equation (8)})
\end{align}
which is the definition of EVSI in Equation (6).

\section{Further details for Case study 1}

\subsection{Further details for calculation of traditional VoI metrics for Case study 1} \label{sec:S1p1}
\noindent In Section 3.1.2 of the main text, we summarised the outputs of traditional VoI calculations for the first case study of the manuscript (protecting frogs from disease outbreaks). Here we explain how these outputs are calculated from application of Equations (1), (2), (5) and (6) to the case study information provided in Section 3.1.1. The description of these calculations here is similar to that presented in \cite{Canessa2015}; however, the present description is also designed to assist with understanding the calculation of the new metrics introduced in the present work (detailed in Section \ref{sec:S1p2}).

First, it is useful to tabulate the system's values $V(a,s)$ for each system state $s \in \{ s_1,s_2 \}$ and chosen management action $a \in \{ a_1,a_2 \}$, as well as the prior probabilities $p(s)$ for each state $s$ (Table~\ref{table:CaseStudy1Values}). All of this information is drawn from the text of Section 3.1.1.

\renewcommand{\arraystretch}{1.3}
\begin{table}[!ht]
\centering
\caption{Values of the system, $V(a,s)$, and prior probabilities $p(s)$, for the frog conservation problem (Case study 1). This table is modified from Table 1 of \cite{Canessa2015}. \\}
\label{table:CaseStudy1Values}
\begin{adjustbox}{center}
\begin{tabular}{|cc|cc|}
\hline
 & & \multicolumn{2}{c|}{States $s$} \\ \cline{3-4}
\multicolumn{2}{|c|}{\textbf{Values} $V(a,s)$} & Disease present & Disease absent \\
& & at new site ($s_1$) & at new site ($s_2$) \\ \hline
\multirow{2}{*}{Actions $a$} & \multicolumn{1}{|c|}{Translocate ($a_1$)} & 55 & 135 \\
& \multicolumn{1}{|c|}{Do nothing ($a_2$)} & 100 & 100  \\ \hline
\multicolumn{2}{|c|}{\textbf{Prior probabilities} $p(s)$} & 0.5 & 0.5 \\ \hline
\end{tabular}
\end{adjustbox}
\end{table}    
\renewcommand{\arraystretch}{1}

From application of Equation (1) to the information shown in Table \ref{table:CaseStudy1Values}, the expected values of the system prior to any measurement, $\text{EV}(a_1)$ and $\text{EV}(a_2)$, if action $a_1$ (translocate) or action $a_2$ (do nothing) is chosen respectively, are:
\begin{align}
\label{eq:EV_a1} \text{EV}(a_1) = \E_{s} [V(a_1,s)] & = \sum_{i=1}^2 {V(a_1,s_i) \cdot p(s_i)}, \nonumber \\
    & = 55 \times 0.5 + 135\times 0.5 = 95, \\ \nonumber \\
\label{eq:EV_a2} \text{EV}(a_2) = \E_{s} [V(a_2,s)] & = \sum_{i=1}^2 {V(a_2,s_i) \cdot p(s_i)}, \nonumber \\
    & = 100 \times 0.5 + 100\times 0.5 = 100.
\end{align}
Combining Equations (2), \eqref{eq:EV_a1} and \eqref{eq:EV_a2}, the expected value of the system if the best management action is carried out prior to any measurement, $\text{EV}_{\text{uncertainty}}$, is:
\begin{equation}
    \text{EV}_{\text{uncertainty}} = \max_a \left\{ \text{EV}(a_1), \text{EV}(a_2) \right\} = \text{EV}(a_2) = 100. \label{eq:EV_uncertainty}
\end{equation}
Therefore, if no measurement is carried out, doing nothing (action $a_2$) is the best action, and this yields an expected total frog population of 100 individuals.

\begin{figure}[ht!]
\centering 
\includegraphics[width=\textwidth]{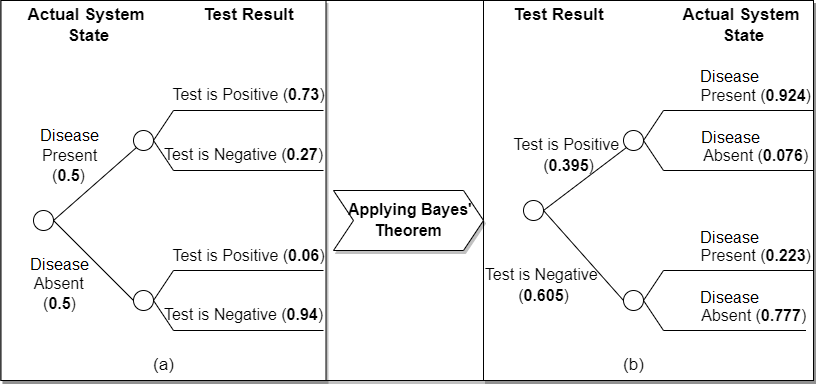}
\caption{\label{fig:CS1DecisionTree} Decision tree for the frog conservation problem. The left side (panel (a)) graphically describes probabilities $p(s)$ of disease presence/absence at the new site and probabilities $p(x|s)$ of test results conditional on disease presence/absence. The right side (panel (b)) graphically describes probabilities $p(x)$ of the test returning positive/negative results and probabilities $p(s|x)$ of disease presence/absence conditional on the test result. Application of Bayes' theorem (Equation \eqref{eq:Bayes}) is used to obtain panel (b) from panel (a).}
\end{figure}

The proposed measurement, which is a test for disease at the new site, has a sensitivity of 0.73 and specificity of 0.94 (Section 3.1.1). If $x_1$ and $x_2$ denote test outcomes that are positive and negative for the disease at the new site, respectively, the information about test sensitivity and specificity can be mathematically represented as:
\begin{equation}
p(x_1|s_1) = 0.73, \qquad p(x_2|s_1) = 0.27, \qquad p(x_1|s_2) = 0.06, \qquad p(x_2|s_2) = 0.94.
\label{eq:p_x_s}
\end{equation}
This information is represented graphically in the left side of Figure \ref{fig:CS1DecisionTree}.

Usage of Equation (5) requires calculation of conditional expectations of the form $\E_{s|x} \{ \cdot \}$, which in turn requires knowledge of conditional probabilities $p(s|x)$. Conditional probabilities $p(s|x)$ can be obtained from conditional probabilities $p(x|s)$ via Bayes' theorem, which for discrete states~$s_i$ and discrete measurement outcomes~$x_j$ is:
\begin{equation}
p(s_i|x_j) = \dfrac{p(x_j|s_i) \, p(s_i)}{p(x_j)}, \qquad \mbox{where} \qquad p(x_j) = \displaystyle\sum_k p(x_j|s_k)p(s_k). \label{eq:Bayes}
\end{equation}
Tedious application of Equation \eqref{eq:Bayes} to Equation \eqref{eq:p_x_s} yields
\begin{align}
&p(x_1) = 0.395, \qquad p(s_1|x_1) = 0.924, \qquad p(s_2|x_1) = 0.076, \nonumber \\
&p(x_2) = 0.605, \qquad p(s_1|x_2) = 0.223, \qquad p(s_2|x_2) = 0.777.
\label{eq:p_s_x}
\end{align}
This information is represented graphically in the right side of Figure \ref{fig:CS1DecisionTree}.

Next, intermediate quantities $\max_a \{ \E_{s|x_1} [ V(a,s) ] \}$ and $\max_a \{ \E_{s|x_2} [ V(a,s) ] \}$, that are required for calculation of $\text{EV}_{\text{less uncertainty}}$, are obtained from the information present in Table \ref{table:CaseStudy1Values} and Equation \eqref{eq:p_s_x}:
\begin{align}
\max_a \{ \E_{s|x_1} [ V(a,s) ] \} &= \max \left\{ \sum_{i=1}^2 V(a_1,s_i) \cdot p(s_i|x_1), \,\, \sum_{i=1}^2 V(a_2,s_i) \cdot p(s_i|x_1) \right\} \nonumber \\
 & = \max \left\{ 55 \times 0.924 + 135 \times 0.076, \,\, 100 \times 0.924 + 100 \times 0.076 \right\} \nonumber \\
 & = 100, \label{eq:max_a_E_s_x1_V_a_s}\\ \nonumber \\
\max_a \{ \E_{s|x_2} [ V(a,s) ] \} &= \max \left\{ \sum_{i=1}^2 V(a_1,s_i) \cdot p(s_i|x_2), \,\, \sum_{i=1}^2 V(a_2,s_i) \cdot p(s_i|x_2) \right\} \nonumber \\
 & = \max \left\{ 55 \times 0.223 + 135 \times 0.777, \,\, 100 \times 0.223 + 100 \times 0.777 \right\} \nonumber \\
 & = 117.2. \label{eq:max_a_E_s_x2_V_a_s}
\end{align}
Then, combining Equations (5) and \eqref{eq:p_s_x}-\eqref{eq:max_a_E_s_x2_V_a_s}, the expected value of the system if the best management action is performed after the test is carried out, $\text{EV}_{\text{less uncertainty}}$, is:
\begin{align}
\text{EV}_{\text{less uncertainty}} = \,\, & \E_{x} [ \max_a \{ \E_{s|x} [ V(a,s) ] \} ] \nonumber \\ 
                                      = \,\, & p(x_1) \, \max_a \{ \E_{s|x_1} [ V(a,s) ] \} + p(x_2) \, \max_a \{ \E_{s|x_2} [ V(a,s) ] \} \nonumber \\
                                      = \,\, & 0.395 \times 100 + 0.605 \times 117.2 \nonumber \\
                                      = \,\, & 110.4. \label{eq:EV_less_uncertainty}
                                       \end{align}
Finally, combining Equations (6), \eqref{eq:EV_uncertainty} and \eqref{eq:EV_less_uncertainty}, the EVSI for this case study is:
\begin{align}
\text{EVSI} &= \text{EV}_{\text{less uncertainty}} - \text{EV}_{\text{uncertainty}} \nonumber \\
			&= 110.4 - 100 \nonumber \\
			&= 10.4. \label{eq:EVSI}
\end{align}
Therefore, the expected value of the test for disease at the new site is an increase in the total frog population of 10.4 individuals.

\subsection{Further details for calculation of new VoI metrics for Case study 1}
\label{sec:S1p2}
\noindent In Section 3.1.3 of the main text, we summarised the outputs of new VoI calculations for the first case study of the manuscript (protecting frogs from disease outbreaks). Here we explain how these outputs are calculated from application of Equations (7)-(11) to the case study information provided in Section 3.1.1. We also here make use of quantities already computed in Section \ref{sec:S1p1} for the present case study.

Combining Equations (7), \eqref{eq:EV_uncertainty}, \eqref{eq:max_a_E_s_x1_V_a_s} and \eqref{eq:max_a_E_s_x2_V_a_s}, the shifts in expected value of the system, $\Delta \text{EV}_{x_1}$ and $\Delta \text{EV}_{x_2}$, if measurement outcome $x_1$ (test positive for disease) or $x_2$ (test negative for disease) are obtained respectively, are:
\begin{align}
\Delta \text{EV}_{x_1} &= \max_a \{ \E_{s|x_1} [V(a,s)] \}  - \text{EV}_{\text{uncertainty}} = 100 - 100 = 0, \\
\Delta \text{EV}_{x_2} &= \max_a \{ \E_{s|x_2} [V(a,s)] \}  - \text{EV}_{\text{uncertainty}} = 117.2 - 100 = 17.2.
\end{align}
Using Equation (8), the optimal action under uncertainty $a^*$ is $a_2$ (do nothing); this can already be seen from Equation \eqref{eq:EV_uncertainty}. Usage of Equation (9) requires the intermediate quantities $\E_{s|x_1} [V(a^*,s)]$ and $\E_{s|x_2} [V(a^*,s)]$, which can themselves be calculated from information present in Table \ref{table:CaseStudy1Values} and Equation \eqref{eq:p_s_x}:
\begin{align}
\E_{s|x_1} [V(a^*,s)] &= \E_{s|x_1} [V(a_2,s)] \nonumber \\
                      &= V(a_2,s_1) \, p(s_1|x_1) + V(a_2,s_2) \, p(s_2|x_1) \nonumber \\
					  &= 100 \times 0.924 + 100 \times 0.076 \nonumber \\
					  &= 100, \label{eq:E_s_x1_V_astar_s} \\ \nonumber \\
\E_{s|x_2} [V(a^*,s)] &= \E_{s|x_2} [V(a_2,s)] \nonumber \\
					  &= V(a_2,s_1) \, p(s_1|x_2) + V(a_2,s_2) \, p(s_2|x_2) \nonumber \\
					  &= 100 \times 0.223 + 100 \times 0.777 \nonumber \\
					  &= 100. \label{eq:E_s_x2_V_star_s}
\end{align}
Then, combining Equations (9), \eqref{eq:max_a_E_s_x1_V_a_s}, \eqref{eq:max_a_E_s_x2_V_a_s}, \eqref{eq:E_s_x1_V_astar_s} and \eqref{eq:E_s_x2_V_star_s}, the values of sample information VSI$_{x_1}$ and VSI$_{x_2}$ dependent on measurement outcomes $x_1$ (test positive for disease) and $x_2$ (test negative for disease) respectively, are:
\begin{align}
\label{eq:VSI_x1} \text{VSI}_{x_1} &= \max_a \{ \E_{s|x_1} [V(a,s)] \}  - \E_{s|x_1} [V(a^*,s)] = 100 - 100 = 0,\\
\label{eq:VSI_x2} \text{VSI}_{x_2} &= \max_a \{ \E_{s|x_2} [V(a,s)] \}  - \E_{s|x_2} [V(a^*,s)] = 117.2 - 100 = 17.2.
\end{align}
Combining Equations (10), \eqref{eq:p_s_x}, \eqref{eq:EVSI}, \eqref{eq:VSI_x1} and \eqref{eq:VSI_x2}, the standard deviation of the value of sample information, $\sigma \text{VSI}$, is given by:
\begin{align}
\sigma \text{VSI} &= \sqrt{\text{Var}_x [\text{VSI}_x]} \nonumber \\
						 &= \sqrt{\E_x \left[ \left(\text{VSI}_x - \E_x \left[ \text{VSI}_x \right] \right)^2 \right]} \nonumber \\
						 &= \sqrt{\E_x \left[ \left(\text{VSI}_x - \text{EVSI} \right)^2 \right]} \nonumber \\
						 &= \sqrt{p(x_1) \, \left(\text{VSI}_{x_1} - \text{EVSI} \right)^2 + p(x_2) \, \left(\text{VSI}_{x_2} - \text{EVSI} \right)^2} \nonumber \\
						 &= \sqrt{0.395 \times (0 - 10.4)^2 + 0.605 \times (17.2-10.4)^2} \nonumber \\
						 &= 8.4.
\end{align}
Finally, since there is a probability of 0.395 that VSI is zero (from Equations \eqref{eq:p_s_x} and \eqref{eq:VSI_x1}), and there is a probability of 0.605 that VSI is 17.2 (from Equations \eqref{eq:p_s_x} and \eqref{eq:VSI_x2}), the risk of the VSI being equal to or less than zero is 0.395. This information, if rewritten using Equation (11), is equivalent to the mathematical statement rVSI$_0 = 0.395$ discussed in Section 3.1.3 of the main text.

\clearpage

\section{Further details for Case study 2}

\subsection{Knowledge of the system prior to any of the proposed trials or measurement actions, for Case study 2}

\renewcommand{\theequation}{S\arabic{section}.\arabic{equation}}
\renewcommand{\thefigure}{S\arabic{section}.\arabic{figure}}
\renewcommand{\thetable}{S\arabic{section}.\arabic{table}}

\renewcommand{\arraystretch}{1.3}
\begin{table}[!htp]
\centering
\caption{Values of the system, $V(a,s)$, and prior probabilities $p(s)$, for the turtle reintroduction problem (Case study 2). Information for this table is obtained from Table 3 of \cite{Canessa2015}. \\}
\label{table:CaseStudy2Values}
\begin{tabular}{|cc|ccc|}
\hline
 & & \multicolumn{3}{c|}{States $s$} \\ \cline{3-5}
\multicolumn{2}{|c|}{\multirow{2}{*}{\textbf{Values} $V(a,s)$}} & \multirow{2}{*}{No post-release} & Post-release & Post-release \\
& & \multirow{2}{*}{effect ($s_1$)} & effect decreases & effect increases \\
& & & with age ($s_2$) & with age ($s_3$) \\ \hline
 & \multicolumn{1}{|c|}{Release 3-} & \multirow{2}{*}{0.689} & \multirow{2}{*}{0.582} & \multirow{2}{*}{0.547} \\
 & \multicolumn{1}{|c|}{year olds ($a_1$)}  & & & \\
\multirow{2}{*}{Actions $a$} & \multicolumn{1}{|c|}{Release 4-} & \multirow{2}{*}{0.729} & \multirow{2}{*}{0.674} &\multirow{2}{*}{0.484} \\
 & \multicolumn{1}{|c|}{year olds ($a_2$)} & & & \\
 & \multicolumn{1}{|c|}{Release 5-} & \multirow{2}{*}{0.745} & \multirow{2}{*}{0.710} & \multirow{2}{*}{0.332} \\
 & \multicolumn{1}{|c|}{year olds ($a_3$)} & & & \\ \hline
\multicolumn{2}{|c|}{\textbf{Prior probabilities} $p(s)$} & 0.4 & 0.2 & 0.4 \\ \hline
\end{tabular}
\end{table}    
\renewcommand{\arraystretch}{1}

\renewcommand{\arraystretch}{1.3}
\begin{table}[!htp]
\centering
\caption{Probability $p_{\text{survival}}(d,s)$ that a single turtle will survive in the wild for one year, dependent on the system state $s$ and the turtle's age at the time of release into the wild (experimental design $d$). Information for this table is obtained from Table 3 of \cite{Canessa2015}. \\}
\label{table:CaseStudy2TrialSurvivalProbabilities}
\begin{tabular}{|cc|ccc|}
\hline
 & & \multicolumn{3}{c|}{States $s$} \\ \cline{3-5}
\multicolumn{2}{|c|}{\textbf{Survival probabilities} $p_{\text{survival}}(d,s)$} & \multirow{2}{*}{No post-release} & Post-release & Post-release \\
\multicolumn{2}{|c|}{\textbf{for a single turtle,}} & \multirow{2}{*}{effect ($s_1$)} & effect decreases & effect increases \\
\multicolumn{2}{|c|}{\textbf{one year in the wild}} & & with age ($s_2$) & with age ($s_3$) \\ \hline
 & \multicolumn{1}{|c|}{Release 3-} & \multirow{2}{*}{0.85} & \multirow{2}{*}{0.72} & \multirow{2}{*}{0.68} \\
Proposed one-year & \multicolumn{1}{|c|}{year olds ($d_1$)}  & & & \\
trial release, & \multicolumn{1}{|c|}{Release 4-} & \multirow{2}{*}{0.90} & \multirow{2}{*}{0.83} &\multirow{2}{*}{0.60} \\
according to & \multicolumn{1}{|c|}{year olds ($d_2$)} & & & \\
experimental design $d$ & \multicolumn{1}{|c|}{Release 5-} & \multirow{2}{*}{0.90} & \multirow{2}{*}{0.86} & \multirow{2}{*}{0.40} \\
 & \multicolumn{1}{|c|}{year olds ($d_3$)} & & & \\ \hline
\end{tabular}
\end{table}
\renewcommand{\arraystretch}{1}

\clearpage
\subsection{Additional results for Case Study 2}

\renewcommand{\arraystretch}{1.3}
\begin{table}[!htp]\footnotesize
\centering
\caption{Probabilities of system state dependent on measurement outcome, $p(x|s)$, of the third proposed experimental design ($d_3$: release 5-year old turtles for one year) in the second case study of improving survival of turtles released from captivity (see Section 3.2 for case study details). Upward-pointing arrows ($\uparrow$) and downward-pointing arrows ($\downarrow$) highlight the increase or decrease, respectively, in the probability of the associated system state $s$ due to measurement outcome $x$, compared with the prior belief of this system state: $p(s_1)=0.4$, $p(s_2)=0.2$ and $p(s_3)=0.4$, as shown in Table \ref{table:CaseStudy2Values}. \\}
\label{table:CaseStudy2_p_x_s_5YearOlds}
\begin{tabular}{ccccc} 
\hline
\multicolumn{5}{c}{ \textbf{Experimental design $d_3$: One-year trial release of 5-year old turtles}} \\ \hline
\textbf{Number of} & \textbf{Probability of} & \multicolumn{3}{c}{ \textbf{ Probability $p(s|x)$ of system state $s$ if outcome $x$ is obtained }} \\ \cline{3-5}
\textbf{surviving} & \textbf{measurement} & \textbf{No post-release} & \textbf{Post-release effect} & \textbf{Post-release effect} \\ 
\textbf{turtles,} $x$ & \textbf{outcome,} $p(x)$ & \textbf{effect ($s_1$)} & \textbf{decreases with age ($s_2$)} & \textbf{increases with age ($s_3$)} \\ \hline
0  & 0.0024 & 0.0000 ($\downarrow$) & 0.0000 ($\downarrow$) & 1.0000 ($\uparrow$)   \\
1  & 0.0161 & 0.0000 ($\downarrow$) & 0.0000 ($\downarrow$) & 1.0000 ($\uparrow$)   \\ 
2  & 0.0484 & 0.0000 ($\downarrow$) & 0.0000 ($\downarrow$) & 1.0000 ($\uparrow$)   \\ 
3  & 0.0860 & 0.0000 ($\downarrow$) & 0.0002 ($\downarrow$) & 0.9998 ($\uparrow$)   \\ 
4  & 0.1006 & 0.0005 ($\downarrow$) & 0.0017 ($\downarrow$) & 0.9977 ($\uparrow$)   \\ 
5  & 0.0821 & 0.0072 ($\downarrow$) & 0.0155 ($\downarrow$) & 0.9772 ($\uparrow$)   \\ 
6  & 0.0556 & 0.0803 ($\downarrow$) & 0.1174 ($\downarrow$) & 0.8022 ($\uparrow$)   \\ 
7  & 0.0629 & 0.3652 ($\downarrow$) & 0.3645 ($\uparrow$)   & 0.2702 ($\downarrow$) \\ 
8  & 0.1345 & 0.5760 ($\uparrow$)   & 0.3924 ($\uparrow$)   & 0.0316 ($\downarrow$) \\ 
9  & 0.2276 & 0.6807 ($\uparrow$)   & 0.3165 ($\uparrow$)   & 0.0028 ($\downarrow$) \\ 
10 & 0.1838 & 0.7589 ($\uparrow$)   & 0.2408 ($\uparrow$)   & 0.0002 ($\downarrow$) \\
\hline
\end{tabular}
\end{table}
\renewcommand{\arraystretch}{1}

\renewcommand{\arraystretch}{1.3}
\begin{table}[!htp]\footnotesize
\centering
\caption{Calculated values for the new VoI metrics that are dependent on measurement outcomes, alongside the probabilities of these outcomes and corresponding flow-on effects to the recommended management action, for the first proposed experimental design ($d_1$: release 3-year old turtles for one year) in the second case study of improving survival of turtles released from captivity (see Section 3.2 for case study details). \\}
\label{table:CaseStudy2FullResults3YearOlds}
\begin{tabular}{ccccc} 
\hline
\multicolumn{5}{c}{ \textbf{Experimental design $d_1$: One-year trial release of 3-year old turtles}} \\ \hline
\textbf{Number of} & \textbf{Shift in expected} & \textbf{Value of sample} & \textbf{Probability of}  & \textbf{Does the recommended }  \\ 
\textbf{surviving} & \textbf{value of the} & \textbf{information,} & \textbf{measurement} & \textbf{action change after the} \\ 
\textbf{turtles,} $x$ & \textbf{system,} $\Delta \text{EV}_x$ & $\text{VSI}_x$ & \textbf{outcome,} $p(x)$ & \textbf{measurement outcome?} \\
\hline
        0 & --0.0689 & 0.0449 & 0.0000 & Yes, release 3-year olds instead \\
        1 & --0.0680 & 0.0416 & 0.0001 & Yes, release 3-year olds instead \\
        2 & --0.0669 & 0.0377 & 0.0011 & Yes, release 3-year olds instead  \\
        3 & --0.0653 & 0.0331 & 0.0064 & Yes, release 3-year olds instead  \\
        4 & --0.0626 & 0.0275 & 0.0252 & Yes, release 3-year olds instead  \\
        5 & --0.0576 & 0.0204 & 0.0694 & Yes, release 3-year olds instead  \\
        6 & --0.0476 & 0.0110 & 0.1391 & Yes, release 3-year olds instead \\
        7 & --0.0279 & 0      & 0.2105 & No, still release 4-year olds  \\
        8 &   0.0135 & 0      & 0.2456 & No, still release 4-year olds  \\
        9 &   0.0544 & 0      & 0.2079 & No, still release 4-year olds  \\
        10 &  0.0854 & 0.0026 & 0.0947 & Yes, release 5-year olds instead \\ \hline
\end{tabular}
\end{table}
\renewcommand{\arraystretch}{1}

\renewcommand{\arraystretch}{1.3}
\begin{table}[!htp]\footnotesize
\centering
\caption{Probabilities of system state dependent on measurement outcome, $p(x|s)$, of the first proposed experimental design ($d_1$: release 3-year old turtles for one year) in the second case study of improving survival of turtles released from captivity (see Section 3.2 for case study details). Upward-pointing arrows ($\uparrow$) and downward-pointing arrows ($\downarrow$) highlight the increase or decrease, respectively, in the probability of the associated system state $s$ due to measurement outcome $x$, compared with the prior belief of this system state: $p(s_1)=0.4$, $p(s_2)=0.2$ and $p(s_3)=0.4$, as shown in Table \ref{table:CaseStudy2Values}. \\}
\label{table:CaseStudy2_p_x_s_3YearOlds}
\begin{tabular}{ccccc} 
\hline
\multicolumn{5}{c}{ \textbf{Experimental design $d_1$: One-year trial release of 3-year old turtles}} \\ \hline
\textbf{Number of} & \textbf{Probability of} & \multicolumn{3}{c}{ \textbf{ Probability $p(s|x)$ of system state $s$ if outcome $x$ is obtained }} \\ \cline{3-5}
\textbf{surviving} & \textbf{measurement} & \textbf{No post-release} & \textbf{Post-release effect} & \textbf{Post-release effect} \\ 
\textbf{turtles,} $x$ & \textbf{outcome,} $p(x)$ & \textbf{effect ($s_1$)} & \textbf{decreases with age ($s_2$)} & \textbf{increases with age ($s_3$)} \\ \hline
0  & 0.0000 & 0.0005 ($\downarrow$) & 0.1162 ($\downarrow$) & 0.8834 ($\uparrow$)   \\
1  & 0.0001 & 0.0012 ($\downarrow$) & 0.1372 ($\downarrow$) & 0.8617 ($\uparrow$)   \\
2  & 0.0011 & 0.0030 ($\downarrow$) & 0.1610 ($\downarrow$) & 0.8359 ($\uparrow$)   \\
3  & 0.0064 & 0.0078 ($\downarrow$) & 0.1875 ($\downarrow$) & 0.8046 ($\uparrow$)   \\
4  & 0.0252 & 0.0198 ($\downarrow$) & 0.2156 ($\uparrow$)   & 0.7646 ($\uparrow$)   \\
5  & 0.0694 & 0.0490 ($\downarrow$) & 0.2420 ($\uparrow$)   & 0.7090 ($\uparrow$)   \\
6  & 0.1391 & 0.1153 ($\downarrow$) & 0.2586 ($\uparrow$)   & 0.6261 ($\uparrow$)   \\
7  & 0.2105 & 0.2467 ($\downarrow$) & 0.2510 ($\uparrow$)   & 0.5023 ($\uparrow$)   \\
8  & 0.2456 & 0.4494 ($\uparrow$)   & 0.2075 ($\uparrow$)   & 0.3431 ($\downarrow$) \\
9  & 0.2079 & 0.6685 ($\uparrow$)   & 0.1401 ($\downarrow$) & 0.1914 ($\downarrow$) \\
10 & 0.0947 & 0.8316 ($\uparrow$)   & 0.0791 ($\downarrow$) & 0.0893 ($\downarrow$) \\
\hline
\end{tabular}
\end{table}
\renewcommand{\arraystretch}{1}

\renewcommand{\arraystretch}{1.3}
\begin{table}[!htp]\footnotesize
\centering
\caption{Calculated values for the new VoI metrics that are dependent on measurement outcomes, alongside the probabilities of these outcomes and corresponding flow-on effects to the recommended management action, for the second proposed experimental design ($d_2$: release 4-year old turtles for one year) in the second case study of improving survival of turtles released from captivity (see Section 3.2 for case study details). \\}
\label{table:CaseStudy2FullResults4YearOlds}
\begin{tabular}{ccccc} 
\hline
\multicolumn{5}{c}{ \textbf{Experimental design $d_2$: One-year trial release of 4-year old turtles}} \\ \hline
\textbf{Number of} & \textbf{Shift in expected} & \textbf{Value of sample} & \textbf{Probability of}  & \textbf{Does the recommended }  \\ 
\textbf{surviving} & \textbf{value of the} & \textbf{information,} & \textbf{measurement} & \textbf{action change after the} \\ 
\textbf{turtles,} $x$ & \textbf{system,} $\Delta \text{EV}_x$ & $\text{VSI}_x$ & \textbf{outcome,} $p(x)$ & \textbf{measurement outcome?} \\
\hline
        0 & --0.0730 & 0.0630 & 0.0000 & Yes, release 3-year olds instead \\
        1 & --0.0730 & 0.0630 & 0.0006 & Yes, release 3-year olds instead \\
        2 & --0.0730 & 0.0628 & 0.0043 & Yes, release 3-year olds instead  \\
        3 & --0.0729 & 0.0625 & 0.0170 & Yes, release 3-year olds instead  \\
        4 & --0.0725 & 0.0612 & 0.0451 & Yes, release 3-year olds instead  \\
        5 & --0.0708 & 0.0570 & 0.0837 & Yes, release 3-year olds instead  \\
        6 & --0.0641 & 0.0438 & 0.1163 & Yes, release 3-year olds instead \\
        7 & --0.0419 & 0.0110 & 0.1410 & Yes, release 3-year olds instead  \\
        8 &   0.0273 & 0      & 0.1844 & No, still release 4-year olds  \\
        9 &   0.0871 & 0.0099 & 0.2347 & Yes, release 5-year olds instead  \\
        10 &  0.1129 & 0.0172 & 0.1729 & Yes, release 5-year olds instead \\ \hline
\end{tabular}
\end{table}
\renewcommand{\arraystretch}{1}

\renewcommand{\arraystretch}{1.3}
\begin{table}[!htp]\footnotesize
\centering
\caption{Probabilities of system state dependent on measurement outcome, $p(x|s)$, of the second proposed experimental design ($d_2$: release 4-year old turtles for one year) in the second case study of improving survival of turtles released from captivity (see Section 3.2 for case study details). Upward-pointing arrows ($\uparrow$) and downward-pointing arrows ($\downarrow$) highlight the increase or decrease, respectively, in the probability of the associated system state $s$ due to measurement outcome $x$, compared with the prior belief of this system state: $p(s_1)=0.4$, $p(s_2)=0.2$ and $p(s_3)=0.4$, as shown in Table \ref{table:CaseStudy2Values}. \\}
\label{table:CaseStudy2_p_x_s_4YearOlds}
\begin{tabular}{ccccc} 
\hline
\multicolumn{5}{c}{ \textbf{Experimental design $d_2$: One-year trial release of 4-year old turtles}} \\ \hline
\textbf{Number of} & \textbf{Probability of} & \multicolumn{3}{c}{ \textbf{ Probability $p(s|x)$ of system state $s$ if outcome $x$ is obtained }} \\ \cline{3-5}
\textbf{surviving} & \textbf{measurement} & \textbf{No post-release} & \textbf{Post-release effect} & \textbf{Post-release effect} \\ 
\textbf{turtles,} $x$ & \textbf{outcome,} $p(x)$ & \textbf{effect ($s_1$)} & \textbf{decreases with age ($s_2$)} & \textbf{increases with age ($s_3$)} \\ \hline
0  & 0.0000 & 0.0000 ($\downarrow$) & 0.0001 ($\downarrow$) & 0.9999 ($\uparrow$)   \\
1  & 0.0006 & 0.0000 ($\downarrow$) & 0.0003 ($\downarrow$) & 0.9997 ($\uparrow$)   \\
2  & 0.0043 & 0.0000 ($\downarrow$) & 0.0010 ($\downarrow$) & 0.9989 ($\uparrow$)   \\ 
3  & 0.0170 & 0.0002 ($\downarrow$) & 0.0033 ($\downarrow$) & 0.9965 ($\uparrow$)   \\
4  & 0.0451 & 0.0012 ($\downarrow$) & 0.0107 ($\downarrow$) & 0.9881 ($\uparrow$)   \\
5  & 0.0837 & 0.0071 ($\downarrow$) & 0.0337 ($\downarrow$) & 0.9592 ($\uparrow$)   \\
6  & 0.1163 & 0.0384 ($\downarrow$) & 0.0986 ($\downarrow$) & 0.8630 ($\uparrow$)   \\
7  & 0.1410 & 0.1629 ($\downarrow$) & 0.2270 ($\uparrow$)   & 0.6101 ($\uparrow$)   \\
8  & 0.1844 & 0.4201 ($\uparrow$)   & 0.3176 ($\uparrow$)   & 0.2623 ($\downarrow$) \\
9  & 0.2347 & 0.6604 ($\uparrow$)   & 0.2709 ($\uparrow$)   & 0.0687 ($\downarrow$) \\
10 & 0.1729 & 0.8066 ($\uparrow$)   & 0.1795 ($\downarrow$) & 0.0140 ($\downarrow$) \\
\hline
\end{tabular}
\end{table}
\renewcommand{\arraystretch}{1}

\clearpage

\bibliographystyle{elsarticle-harv} 
\bibliography{Bibliography}